\documentstyle[aps,preprint]{revtex}
\begin{document}
\title{Structure of Proton}
\author{Fayyazuddin}
\address{National Center for Physics\\
Quaid-e-Azam University\\
Islamabad 45320\\
Pakistan}
\maketitle

\begin{abstract}
Electron--proton scattering in elastic and highly inelastic region is
reviewed in a unified approach. The importance of parity--violating
scattering due to electro--weak interference in probing the structure of
proton is emphasized. The importance of longitudnal spin--spin asymmetry as
well as parity violating longitudnal asymmetry to extract the structure
functions of proton in both regions are discussed. The recoil polarization
of proton in the elastic scattering is also discussed.

Preprint No. NCP-QAU/0306-17
\end{abstract}

\newpage\ 

\section{Introduction}

Lepton--proton (nucleon) scattering is a very direct means of probing
nucleon structure. In particular electron--nucleon scattering serves to
produce a virtual photon of space like four momentum which probes nucleon
structure in a very clean way. Elastic scattering experiments have been
carried out extensivly and we now have a fairly detailed knowledge of
nucleon form factors as functions of virtual photon mass $\left( Q^2\right) $%
. In these experiments, the nucleons recoils elastically, the photon
interacts with the nucleon constituents in a coherent manner and these form
factors are related roughly to average shape of a nucleon. On the other
hand, the inelastic total scattering cross-section is described by the
structure functions $W_2$ and $W_1$ that depend upon both the photon energy $%
\left( \nu \right) $ and the photon mass $\left( Q^2\right) $. In the
inelstic processes the photon interacts in an incoherent manner and it
probes, roughly, the instantaneous construction of the nucleon rather than
the average shape found in the elastic scattering experiments. Thus the
study of inelastic lepton--hadron scattering at high energies and large
momentum transfers (deep inelstic scattering) may give information about the
structure pertaining to any fundamental constituents of a nucleon. In the
high energy inelastic experiments, the quantities measured are the momenta
and scattering angles of electrons. The cross--section thus measured
represents a sum over all the final hadronic states possible for a definite
laboratory energy and hence for an in variant mass of the hadrons. The
elastic scattering can be regarded as a special case of inelastic scattering
in which the invariant hadronic mass is replaced by the mass $\left(
m\right) $ of the nucleon and the structure functions $W_2\left( \nu
,q^2\right) $ and $W_1\left( \nu ,q^2\right) $ are then given by the two
form factors $F_2\left( q^2\right) $ and $F_1\left( q^2\right) $ multiplied
by the $\delta $--function $\delta \left( \nu -q^2/2m\right) $.

To further probe the structure of the proton, it is important to investigate
the spin dependent structure functions. In fact, the European Muon
Collaboration (EMC) measurement of polarized muon-proton deep inelastic
scattering gave the first indication that strange content of the proton is
not zero. Apart from this direct evidence; there were also indications (see
below) that simple picture of a proton having only non-strange valence $u$
and $d$ quarks is not tenable. Instead of using both polarized lepton beams
and polarized proton targets; one can dispense with either polarized proton
targets or polarized lepton beams. The interference between the photon and $%
Z $-boson exchange can supply the missing polarization content. Thus the
parity violating scattering of polarized (unpolarized) electron beam on
unpolarized (polarized) proton target can give information about the matrix
elements $\left\langle p\left| \bar{q}\Gamma _\mu q\right| p\right\rangle $, 
$\Gamma _\mu =i\gamma _\mu ,$ $i\gamma _\mu \gamma _5;$ $q=u,d,s$. In
particular parity violating elastic scattering would give direct information
about the axial and vector form factors $G_A^Z$, $F_1^Z$ and $F_2^Z$ due to $%
Z$-exchange. These form factors involve both isosinglet and octet parts.
While the octet form factors can be extracted from the $\beta $-decay of
hyperon and electromagnetic form factors; the singlet form factors have to
be extracted from the direct measurement of $G_A^Z$, and $F_2^Z$ from the
parity violating electron scattering.

The purpose of this article is to review the structure of proton. We follow
a general approach which is applicable to both elastic and deep inelastic
scattering.

\section{Electron scattering on unpolarized nucleon}

Lord Rutherford was the first physicst to use scattering experiments to
probe the structure of the matter. He used scattering of $\alpha $%
--particles on atoms. He found that his experimental results are compatible
with a point like positive charge inside the atom which he called atomic
nucleus--a discovery of tremendous importance.

We now consider the scattering of electrons on nucleons (see Fig. \ref{fig1}%
) 
\begin{eqnarray*}
k &\equiv &\left( {\bf k},iE\right) ,\text{ }k^{\prime }\equiv \left( {\bf k}%
^{\prime },iE^{\prime }\right) \\
p &\equiv &\left( {\bf 0},im\right) \\
q &=&k-k^{\prime },\text{ }\nu =E-E^{\prime } \\
q^2 &\approx &2EE^{\prime }\left( 1-\cos \theta \right) ,\text{ }m_e\approx
0,\,\,\,\,\,\,\,Q^2=-q^2 \\
\text{For elastic scattering \thinspace \thinspace \thinspace \thinspace }%
p_X &=&p^{\prime } \\
p^{\prime 2} &=&-m^2
\end{eqnarray*}

The differential cross--section is proportional \cite{MIPP1992} to 
\begin{equation}
L_{\mu \nu }J_{\mu \nu }  \label{01}
\end{equation}
where $L_{\mu \nu }$, $J_{\mu \nu }$ are given by 
\begin{eqnarray}
L_{\mu \nu } &=&\frac 1{2m_e^2}\left[ k_\mu k_\nu ^{\prime }+k_\mu k_\nu
^{\prime }-\delta _{\mu \nu }k\cdot k^{\prime }\right]  \label{02} \\
J_{\mu \nu } &=&2\pi \left[ \frac 1m\left( p_\mu -\frac{p\cdot q}{q^2}q_\mu
\right) \left( p_\nu -\frac{p\cdot q}{q^2}q_\nu \right) W_2\left( \nu
,q^2\right) \right.  \nonumber \\
&&+\left. \left( \delta _{\mu \nu }-\frac{q_\mu q_\nu }{q^2}\right)
W_1\left( \nu ,q^2\right) \right]  \label{03}
\end{eqnarray}
Note that $L_{\mu \nu }$ is the leptonic part which we get after summing
over the final electron spin and taking the spin average of initial
electron. Leptonic part is completely known. The hadronic part is described
in terms of two structure functions $W_2$ and $W_1$ after taking the spin
average of the target nucleon and summing over all the quantum numbers of $%
p_X$.

The differential cross-section is given by 
\begin{equation}
\frac{d^2\sigma }{dq^2d\nu }=\left( \frac{d\sigma }{dq^2}\right) _{\text{Mott%
}}\left[ W_2\left( \nu ,q^2\right) +2\tan ^2\frac \theta 2W_1\left( \nu
,q^2\right) \right]  \label{04}
\end{equation}
where 
\begin{equation}
\left( \frac{d\sigma }{dq^2}\right) _{\text{Mott}}=\frac{4\pi \alpha ^2}{q^4}%
\frac{E^{\prime }}E\cos ^2\frac \theta 2  \label{05}
\end{equation}
give the differential cross-section for the scattering of electron on
spinless structureless proton. The presence of the structure functions in
Eq. (\ref{04}) indicates that proton is not a point particle.

For elastic scattering $\nu =q^2/2m$. For this case the structure functions
are given by 
\begin{eqnarray}
W_2 &=&\left[ F_1^2\left( q^2\right) +\tau F_2^2\left( q^2\right) \right]
\delta \left( \nu -\frac{q^2}{2m}\right)  \nonumber \\
W_1 &=&\left[ F_1\left( q^2\right) +F_2\left( q^2\right) \right] ^2\delta
\left( \nu -\frac{q^2}{2m}\right)  \label{06}
\end{eqnarray}
where $\tau =q^2/4m^2$. Thus from (\ref{04}), we get \cite{PR79-615} 
\begin{equation}
\frac{d\sigma }{dq^2}=\left( \frac{d\sigma }{dq^2}\right) _{\text{Mott}%
}\left\{ \frac{G_E^2\left( q^2\right) +\tau G_M^2\left( q^2\right) }{1+\tau }%
+2\tan ^2\frac \theta 2G_M^2\left( q^2\right) \right\}  \label{07}
\end{equation}
where 
\begin{eqnarray}
G_E\left( q^2\right) &=&F_1\left( q^2\right) -\tau F_2\left( q^2\right) 
\nonumber \\
G_M\left( q^2\right) &=&F_1\left( q^2\right) +F_2\left( q^2\right)
\label{08}
\end{eqnarray}
The Pauli form factors $F_1$ and $F_2$ are normalized as follows: 
\begin{eqnarray}
F_1^p\left( 0\right) &=&1,\text{ }F_2^p\left( 0\right) =\kappa
_p\,\,\,\,\,\,\,\,\,\,\,\,\,\,\,\,\,\,\,\,\,\kappa _p=1.792  \nonumber \\
F_1^n\left( 0\right) &=&0,\text{ }F_2^n\left( 0\right) =\kappa
_n\,\,\,\,\,\,\,\,\,\,\,\,\,\,\,\,\,\,\,\,\,\kappa _n=-1.913  \label{09}
\end{eqnarray}
$\kappa _p$ and $\kappa _n$ are the anomalous magnetic moments of the proton
and the neutron respectively. Experimental data is analysed in terms of
Sachs form factors $G_E\left( q^2\right) $ and $G_M\left( q^2\right) $ which
are normalized as follows: 
\begin{eqnarray}
G_E^p\left( 0\right) &=&1,\text{ }G_M^p=\mu _p=2.792  \nonumber \\
G_E^n\left( 0\right) &=&0,\text{ }G_M^n=\mu _n=-1.913  \label{10}
\end{eqnarray}
>From the fig. 1 and Eq. (\ref{04}), it is easy to see that the structure
functions $W_1$ and $W_2$ are related to the absorptive part for the forward
Compton scattering for virtual photon. In fact using optical theorem one
finds 
\begin{equation}
\left( 1+\frac{\nu ^2}{q^2}\right) \frac{W_2}{W_1}-1=\frac{\sigma _L}{\sigma
_T},  \label{11}
\end{equation}
where $\sigma _L$ and $\sigma _T$ are the longitudnal and transverse total
Compton scattering cross--section respectively. For the elastic scattering,
Eq. (\ref{11}) gives 
\begin{equation}
\frac{\sigma _L}{\sigma _T}=\frac{G_E^2}{\tau G_M^2}  \label{12}
\end{equation}
The virtual longitudnal polarization $\epsilon $ is given by 
\begin{equation}
\epsilon =\frac 1{1+2\left( 1+\frac{\nu ^2}{q^2}\right) \tan ^2\frac \theta 2%
}=\frac 1{1+2\left( 1+\tau \right) \tan ^2\frac \theta 2}  \label{13}
\end{equation}
The Rosenbluth scattering cross-section given in Eq. (\ref{07}) can be
written as follows \cite{PR79-615}: 
\begin{equation}
\frac{d\sigma }{dq^2}=\left( \frac{d\sigma }{dq^2}\right) _{\text{Mott}}%
\frac 1{\epsilon \left( 1+\tau \right) }\left[ \tau G_M^2+\epsilon
G_E^2\right]  \label{14}
\end{equation}
It is clear from Eq. (\ref{14}) that $G_M$ can be extracted from the
scattering cross--section at $\epsilon =0$, while $G_E$ is extracted from
the $\epsilon $--dependance. $\epsilon $ can be varied at fixed photon
energy and momentum $\left( {\bf q},i\nu \right) $ by varying the electron
energy and scattering angle. A global analysis of data indicated that $G_E$
and $G_M$ follows the dipole form, $G_D=\left( 1+q^2/0.71\right) ^{-2}$; $%
G_M $ with great accuracy while $G_E$ with less precisely as to extract $G_E$
at high $q^2$ is less accurate; because of $\epsilon /\tau $ weighting of
the electric term relative to the magnetic term. For, $q^2<1$ GeV$^2$, one
finds 
\[
\frac{G_M}{\mu _pG_D}\approx \frac{G_E}{G_D}\approx 1 
\]
However, recent measurements of recoil polarization of proton in \cite
{PRL84-1398,PRL88-092301} elastic scattering show that ratio $G_M/G_E$
systematically decreases as $q^2$ increases from 0.5 to 3.5 GeV$^2$,
indicating for the first time a definite difference in the spatial
distribution of charge and magnetization currents in the proton. If
confirmed, it will indicate the breakdown of universality of electric and
magnetic distribution at high $q^2$ in a proton.

\section{Strangeness in the protron}

The simple picture of the protron having only non--strange valence u and d
quarks has been questioned In fact it was shown for quite some time that the 
$\Sigma $ term in pion--nucleon scattering \cite{CTP1466} which is given by 
\begin{equation}
\sum\limits_{\pi N}\left( 0\right) =\frac 12\left\langle p\left| \left[
F_{1-i2}^5,\left[ F_{1+i2}^5,H_M\right] \right] \right| p\right\rangle
\label{15}
\end{equation}
where $F_i^5$ are the axial--vector charges and $H_M$ is the
chiral--symmetry breaking Hamiltonian in QCD: 
\begin{eqnarray}
H_M &=&m_u\bar{u}u+m_d\bar{d}d+m_s\bar{s}s  \nonumber \\
&=&\sqrt{6}m_0S_0+\frac 2{\sqrt{3}}\left( \bar{m}-m_s\right) S_8+\left(
m_u-m_d\right) S_3  \label{16}
\end{eqnarray}
where 
\begin{eqnarray}
m_0 &=&\frac{m_u+m_d+m_s}{\sqrt{3}},\text{ \thinspace \thinspace \thinspace
\thinspace \thinspace \thinspace \thinspace \thinspace \thinspace \thinspace
\thinspace \thinspace \thinspace }\bar{m}=\frac{m_u+m_d}2  \nonumber \\
S_0 &=&\frac{\bar{u}u+\bar{d}d+\bar{s}s}{\sqrt{6}},\,\,\,\,\,S_8=\frac{\bar{u%
}u+\bar{d}d-2\bar{s}s}{2\sqrt{3}},\,\,\,\,\,\,S_3=\frac{\bar{u}u-\bar{d}d}2
\label{17}
\end{eqnarray}
>From above equations, one can write 
\begin{eqnarray}
\sum\limits_{\pi N}\left( 0\right) &=&2\bar{m}\left\langle p\left| \frac 1{%
\sqrt{3}}\left( S_8+\sqrt{2}S_0\right) \right| p\right\rangle  \nonumber \\
&=&\bar{m}\left\langle p\left| \bar{u}u+\bar{d}d\right| p\right\rangle
\label{18}
\end{eqnarray}
Now $\left\langle p\left| S_8\right| p\right\rangle $ is entirely determined
from the Gell--Mann--Okubo mass splitting: 
\begin{equation}
\left\langle p\left| S_8\right| p\right\rangle =\frac{\sqrt{3}}2\frac{m_\Xi
-m_\Lambda }{m_s-\bar{m}}  \label{19}
\end{equation}
Above pattern of symmetry breaking implies ($\lambda _0=\sqrt{2}\lambda _8$) 
\begin{equation}
\left\langle p\left| S_0\right| p\right\rangle =\sqrt{2}\left\langle p\left|
S_8\right| p\right\rangle  \label{20}
\end{equation}
Thus one gets 
\begin{equation}
\sum\limits_{\pi N}\left( 0\right) =\frac{3\left( m_\Xi -m_\Lambda \right) 
\bar{m}}{m_s-\bar{m}}  \label{21}
\end{equation}
Using $\bar{m}/\left( m_s-\bar{m}\right) =0.4$ \cite{PRD13-2161}, and
experimental masses for $m_\Xi $ and $m_\Lambda $, one obtains 
\begin{equation}
\sum\limits_{\pi N}\left( 0\right) \approx 25\text{ MeV}  \label{22}
\end{equation}
which is about half the value of $\sum\limits_{\pi N}\left( 0\right) $
extracted from low--energy pion--nucleon scattering: namely \cite{CTP1466} 
\begin{equation}
\sum\limits_{\pi N}\left( 0\right) =51\pm 5\text{ MeV}  \label{23}
\end{equation}
Experimental value of $\sum\limits_{\pi N}\left( 0\right) $ implies that Eq.(%
\ref{20}) is not valid and has to be modified. Let us write 
\begin{mathletters}
\begin{equation}
\sqrt{2}\left\langle p\left| S_0\right| p\right\rangle =2\left( 1+C_s\right)
\left\langle p\left| S_8\right| p\right\rangle  \label{24a}
\end{equation}
so that 
\begin{equation}
\left\langle p\left| \bar{s}s\right| p\right\rangle =\frac 23C_s\frac 1{1+%
\frac 23C_s}\left\langle p\left| \frac 12\left( \bar{u}u+\bar{d}d\right)
\right| p\right\rangle  \label{24b}
\end{equation}
Then Eq. (\ref{18}) is modified as follows 
\end{mathletters}
\begin{eqnarray}
\sum\limits_{\pi N}\left( 0\right) &=&\frac 2{\sqrt{3}}\bar{m}\left(
3+2C_s\right) \left\langle p\left| S_8\right| p\right\rangle  \nonumber \\
&=&\frac{\left( m_\Xi -m_\Lambda \right) \bar{m}}{m_s-\bar{m}}\left(
3+2C_s\right)  \label{25}
\end{eqnarray}

Comparing it with the experimental value (Eq. \ref{23}), one gets 
\begin{equation}
C_s=\frac 32  \label{26}
\end{equation}

Now in the valence quark model if the proton primarily consists u and d
quarks one must have \cite{PR175-2195} 
\begin{equation}
\left\langle p\left| \bar{s}s\right| p\right\rangle \ll \left\langle p\left| 
\frac 12\left( \bar{u}u+\bar{d}d\right) \right| p\right\rangle  \label{27}
\end{equation}
But Eq. (24) and (\ref{26}) implies that 
\begin{equation}
\left\langle p\left| \bar{s}s\right| p\right\rangle =\frac 12\left\langle
p\left| \frac 12\left( \bar{u}u+\bar{d}d\right) \right| p\right\rangle
\label{28}
\end{equation}
that is proton has about 50\% probability of containing $\bar{s}s$ pairs.
The same conclusion viz that strange content of a proton is not negligible
was derived in Ref. \cite{PRD38-944}. We considered the effect of isospin
violating part (i.e. the third term) of the Hamiltonian (\ref{16}) on the
pion nucleon vertex function viz 
\begin{equation}
\Gamma =\left\langle p\pi ^0\left| H_{QCD}^{\Delta I=1}\right| p\right\rangle
\label{29}
\end{equation}
In the soft pion limit, one gets \cite{PRD38-944} 
\begin{equation}
\Gamma =-\frac{m_d-m_u}{F_\pi }\frac 2{\sqrt{3}}\left\langle p\left| \left(
P_8+\sqrt{2}P_0\right) \right| p\right\rangle  \label{30}
\end{equation}

Note that the same combination of the pseudoscalar densities enters as that
for the scalar desities in the $\sum\limits_{\pi N}$. Using PCAC and flavor
SU(3), we get 
\begin{equation}
\frac{\delta g}{g_\pi }=\left( m_d-m_u\right) \frac{3F/D-1}{F/D+1}\frac{1+%
\frac 23C_p}{\frac 23\left[ \left( \bar{m}+m_s\right) +2\left( 1+C_p\right)
\left( \bar{m}-m_s\right) \right] }  \label{31}
\end{equation}

Now since $\delta g/g_\pi $ has no pion pole, it should vanish in the chiral
limit $\left( m_d,m_u\rightarrow 0\right) $ However, for $C_p=0$, Eq. (\ref
{31}) reduces to 
\begin{equation}
\frac{\delta g}{g_\pi }=\sqrt{3}\frac{m_d-m_u}{m_d+m_u}\left[ \frac 1{\sqrt{3%
}}\frac{3F/D-1}{F/D+1}\right]  \label{32a}
\end{equation}
which does not vanish in the chiral limit. Hence $C_P$ cannot be zero. i.e.
strange content of the proton is not negligible. This conclusion derived
without any experimental input but purely from the validity of chiral SU(2)$%
\times $SU(2) symmetry. It is intersting to see \cite{PRD38-944} that in the
chiral SU(3)$\times $SU(3) limit 
\begin{equation}
C_P\approx -1  \label{32b}
\end{equation}

Finally, since, in the non--relativistic quark model

\begin{equation}
\bar{q}i\gamma _5q\sim \bar{q}\frac{{\bf \sigma }\cdot \left( {\bf p}%
^{\prime }-{\bf p}\right) }{2m_q}q  \label{33}
\end{equation}
where $m_q$ is the constituent--quark mass, the value $C_P\approx -1$
implies 
\begin{equation}
\left\langle p\left| -\bar{s}i\gamma _5s\right| p\right\rangle \approx
2\left\langle p\left| \frac 12\left( \bar{u}i\gamma _5u+\bar{d}i\gamma
_5d\right) \right| p\right\rangle  \label{34}
\end{equation}

This indicates that the strange quarks are polarized opposite to the
proton's spin while the up and down quarks are polarized along the proton
spin

Additional piece of evidence that proton has strange--quark content comes
from the European Muon Collaboration (EMC) data \cite{PR162C45} for the
polarized electroproduction structure function. These experiments involve
the scattering of polarized electron beam of the polarized proton target in
the deep inelastic region. Since in the deep inelastic region, one gets
information about the elementary constituents of the proton; these
experiments directly probe the spin content of protons in terms of quark
spin. We define the quark contribution to proton spin: \cite
{MIPP1992,MIPP2000} 
\begin{equation}
\left\langle p\left| -\bar{q}i\gamma _\mu \gamma _5q\right| p\right\rangle
=\Delta \tilde{q}\left( -S_\mu \right) ,\;\;\;\;\;\;\;\;\;\;\;\;\;q=u,d,s
\label{35}
\end{equation}
where $S_\mu =\bar{\psi}i\gamma _\mu \gamma _5\psi $ is the spin of the
proton. Taking into account, the gluon contribution to the proton spin. 
\begin{equation}
\Delta \tilde{q}=\Delta q-\frac{\alpha _s}{2\pi }\Delta G_q  \label{36}
\end{equation}
However, this sepration is not unambigous. $\Delta \tilde{q}$ are related to
the axial vector coupling constants $g_{A\text{, }}g_A^8$ and $g_A^0$ as
follows

\begin{eqnarray}
\Delta \tilde{u}-\Delta \tilde{d} &=&g_A=1.2670\pm 0.0035  \nonumber \\
\Delta \tilde{u}+\Delta \tilde{d}-2\Delta \tilde{s} &=&g_A^8=3F-D  \nonumber
\\
F &=&0.463\pm 0.023  \nonumber \\
D &=&0.803\pm 0.040  \label{37}
\end{eqnarray}
These experimental values are obtained from axial vector coupling in $\beta $%
--decay of hyperons. However for the singlet constant 
\[
g_A^0=\Delta \tilde{u}+\Delta \tilde{d}+\Delta \tilde{s} 
\]
one cannot get information from $\beta $--decay; because the known hyperons
belong to Octet of SU(3). EMC data give \cite{PR162C45,PRD41-3517}:

\begin{eqnarray}
\Delta \tilde{u} &=&0.78\pm 0.07  \nonumber \\
\Delta \tilde{d} &=&-0.48\pm 0.08  \nonumber \\
\Delta \tilde{s} &=&-0.14\pm 0.07  \label{38}
\end{eqnarray}
so that 
\begin{equation}
\Delta \tilde{u}+\Delta \tilde{d}+\Delta \tilde{s}=0.16\pm 0.22  \label{39}
\end{equation}
which is consistent with zero. In other words 
\begin{equation}
g_A^0\equiv \Delta \tilde{\Sigma}=\Delta \Sigma -\frac{3\alpha _s}{2\pi }%
\Delta \tilde{G}=0.16\pm 0.22  \label{40}
\end{equation}
where $\Delta \Sigma =\Delta u+\Delta d+\Delta s$ is the quark contribution
to the spin of the proton and $\Delta \tilde{G}$ is the singlet part of $%
\Delta G$. Various estimates of $\Delta \Sigma $ indicate that $\Delta
\Sigma =0$, which implies that $\frac{\alpha s}{2\pi }\left( -\Delta \tilde{G%
}\right) =0.05\pm 0.07$. Thus one can say that the quarks do not contribute
to the spin of the proton (this is known as spin crises for the proton)
implying in view of angular momentum sum rule

\begin{equation}
\frac 12=\Delta \Sigma +\Delta \tilde{G}+L_z  \label{41}
\end{equation}
that its spin is carried out by gluons and/or orbital angular momentum of
its constituents. $\Delta \Sigma \simeq 0$ is in complete disagreement with
the naive quark model (NQM) result which predicts $\Delta \Sigma =1$. Thus
it is very important to measure both $g_A^0$ and $F_2^0\left( 0\right) $
[the SU(3), singlet anomalous magnetic moment of the proton] experimentally
in order to determine the flavor and spin of content of the proton.

Since straight forward interpretation of $g_A^0$ in terms of the quark
contribution to the proton spin is not possible due to the anomely in the
iso--singlet axial vector current; (cf. Eq. (\ref{40})); it is important to
determine both $g_A^0$ and $F_2^0\left( 0\right) $ directly from experiment.
While $g_A^0$ can be directly determined by neutrino--proton elastic
scattering \cite{PRD35-785}, but experimentally $F_2^0\left( 0\right) $ can
only be determined by electron proton scattering. In refrence \cite
{PLB219-140}, it was suggested that parity violating polarized electron
scattering on unpolarized protons (nucleons) can give information about $%
g_A^0$ and $F_2^0\left( 0\right) $ and hence about the matrix elements $%
\left\langle p\left| \bar{s}\Gamma _\mu s\right| p\right\rangle $, $\Gamma
_\mu =i\gamma _\mu $, $i\gamma _\mu \gamma _5$. In 1990, it was suggested in
Ref. \cite{PRD42-794} that elastic parity violating unpolarized electron
scattering on polarized protons can give information about iso-singlet
axial--vector $\left( g_A^0\right) $ and vector $\left( F_2^0\right) $ form
factors. These experiments thus can throw some light on the strange content
of the proton. For a review and recent work see Ref. \cite{9811055}.

\section{Parity--violating electron scattering and proton form factors}

It is well known that to get information for spin--dependent structure
functions \cite{PRD14-1467,PRD9-1444}; one needs polarized electron beam and
polarized proton target. However, for the electron--proton scattering, the
interference between photon and $Z$--boson exchange diagrams (see Fig. 1;
for $Z$--boson exchange replace $\gamma \rightarrow Z$) can also give us
information by scattering of polarized (unpolarized) electrons on
unpolarized (polarized) protons.

In electroweak theory (see for example Refs: \cite{MIPP1992} and \cite
{PRD41-3517}), the electromagnetic and weak neutral currents coupled to
photon and Z--boson respectively are given by for the quarks 
\begin{eqnarray}
J_\mu ^{\text{e.m.}} &=&i\bar{q}\gamma _\mu Qq=i\bar{q}\gamma _\mu \frac 12%
\left[ \lambda _3+\frac 1{\sqrt{3}}\lambda _8\right] q  \nonumber \\
J_\mu ^Z &=&i\bar{q}\gamma _\mu \left( 1+\gamma _5\right) \frac 12\left[
\lambda _3+\frac 1{\sqrt{3}}\lambda _8-\frac 1{\sqrt{6}}\lambda _0\right]
q-2x_WJ_\mu ^{\text{e.m.}}  \label{42}
\end{eqnarray}
For electron, we have 
\begin{eqnarray}
J_\mu ^{\text{e.m.}} &=&i\bar{e}\gamma _\mu e  \nonumber \\
J_\mu ^Z &=&\frac i2\left[ g_V\bar{e}\gamma _\mu e+g_A\bar{e}\gamma _\mu
\gamma _5e\right]  \label{43} \\
g_V &=&-\frac 12+2x_W=-\frac 12\left( 1-4x_W\right) =\frac 12v_e  \nonumber
\\
g_A &=&-\frac 12=\frac 12a_e  \nonumber \\
x_W &=&\sin ^2\theta _W  \label{44}
\end{eqnarray}
The vector and axial--vector form factors of the proton are defined as 
\begin{equation}
\left( 2\pi \right) ^3\left( \frac{p_0p_0^{\prime }}{m^2}\right)
^{1/2}\left\langle p^{\prime }\left| J_\mu ^Z\right| p\right\rangle =\bar{u}%
\left( p^{\prime }\right) i\left[ F_1^Z\gamma _\mu -\frac{F_2^Z}{2m}\sigma
_{\mu \lambda }q^\lambda +G_A^Z\gamma _\mu \gamma _5\right]  \label{45}
\end{equation}
For $J_\mu ^{\text{e.m.}}$, $F_1^Z\rightarrow F_1^\gamma $, $%
F_2^Z\rightarrow F_2^\gamma $, $G_A^\gamma =0$. Note form factors are
functions of $q^2$

Thus we have 
\begin{eqnarray}
F_{1,2}^{\gamma p}\left( 0\right) &=&F_{1,2}^3\left( 0\right) +\frac 1{\sqrt{%
3}}F_{1,2}^8\left( 0\right)  \nonumber \\
F_{1,2}^{Zp}\left( 0\right) &=&\left( 1-2x_w\right) \left[ F_{1,2}^3\left(
0\right) +\frac 1{\sqrt{3}}F_{1,2}^8\left( 0\right) \right] -\frac 1{\sqrt{6}%
}F_{1,2}^0\left( 0\right)  \nonumber \\
G_A^{Zp}\left( 0\right) &=&G_A^3\left( 0\right) +\frac 1{\sqrt{3}}%
G_A^8\left( 0\right) -\frac 1{\sqrt{6}}G_A^0\left( 0\right)  \label{46}
\end{eqnarray}
Since 
\begin{equation}
F_1^{\gamma p}\left( 0\right) =1,\;\;\;F_2^{\gamma p}\left( 0\right) =\kappa
_p,\;\;\;F_1^{\gamma n}\left( 0\right) =0,\,\,\,\,\,\,\,F_2^{\gamma n}\left(
0\right) =\kappa _n  \label{47}
\end{equation}
we get 
\begin{eqnarray}
F_1^3\left( 0\right) &=&\frac 12,\;\;\;F_1^8\left( 0\right) =\frac{\sqrt{3}}2%
,\;\;\;F_1^0\left( 0\right) =\sqrt{\frac 32}  \nonumber \\
F_2^3\left( 0\right) &=&\frac 12\left( \kappa _p-\kappa _n\right)
,\;\;\;F_2^8\left( 0\right) =\frac{\sqrt{3}}2\left( \kappa _p+\kappa
_n\right) ,\;\;\;F_2^0\left( 0\right) =\sqrt{\frac 32}\kappa _0  \label{48}
\end{eqnarray}
Note that $F_2^0\left( 0\right) $ is not fixed i.e. isosinglet anamalous
magnetic moment is not known experimentally. In terms of $u$, $d$, and $s$
quarks: 
\begin{eqnarray}
F_{1,2}^3\left( 0\right) &=&\frac 12\left( F_{1,2}^u\left( 0\right)
-F_{1,2}^d\left( 0\right) \right)  \nonumber \\
F_{1,2}^8\left( 0\right) &=&\frac 12\frac 1{\sqrt{3}}\left( F_{1,2}^u\left(
0\right) +F_{1,2}^d\left( 0\right) -2F_{1,2}^s\left( 0\right) \right) 
\nonumber \\
F_{1,2}^0\left( 0\right) &=&\frac 12\sqrt{\frac 23}\left( F_{1,2}^u\left(
0\right) +F_{1,2}^d\left( 0\right) +F_{1,2}^s\left( 0\right) \right)
\label{49}
\end{eqnarray}
Thus we get 
\begin{eqnarray}
F_1^u\left( 0\right) &=&2,\;\;\;F_1^d\left( 0\right) =1,\;\;\;F_1^s\left(
0\right) =0  \nonumber \\
F_2^u\left( 0\right) &=&\kappa _p+\kappa _0,\;\;\;F_2^d\left( 0\right)
=\kappa _n+\kappa _0,\;\;\;F_2^s\left( 0\right) =\kappa _0-\left( \kappa
_p+\kappa _n\right)  \label{50}
\end{eqnarray}
For $Z$-exchange we have 
\begin{eqnarray}
F_1^{Zp}\left( 0\right) &=&\frac 12\left( 1-4x_w\right)
,\,\,\,\,\,\,\,\,\,\,\,F_2^{Zp}\left( 0\right) =\left( 1-2x_w\right) \kappa
_p-\frac 12\kappa _0  \label{51} \\
G_A^{Zp}\left( 0\right) &=&\frac 12\left[ \left( F-D\right) +\frac 13\left(
3F-D\right) -\frac 13g_A^0\right]  \label{55}
\end{eqnarray}
where in writing Eq. (\ref{55}), we have used Eq. (\ref{46}) 
\begin{eqnarray}
G_A^3\left( 0\right) &=&\frac 12\left( G_A^u\left( 0\right) -G_A^d\left(
0\right) \right) =\frac 12g_A=\frac 12\left( F+D\right)  \nonumber \\
G_A^8\left( 0\right) &=&\frac 12\frac 1{\sqrt{3}}\left( G_A^u\left( 0\right)
+G_A^d\left( 0\right) -2G_A^s\left( 0\right) \right) =\frac 12\frac 1{\sqrt{3%
}}\left( 3F-D\right)  \nonumber \\
G_A^0\left( 0\right) &=&\frac 12\sqrt{\frac 23}\left( G_A^u\left( 0\right)
+G_A^d\left( 0\right) +G_A^s\left( 0\right) \right) =\frac 12\sqrt{\frac 23}%
g_A^0  \label{56}
\end{eqnarray}

Further we note that 
\begin{eqnarray}
G_A^u\left( 0\right) &=&F+\frac 13D+\frac 13g_A^0  \nonumber \\
G_A^d\left( 0\right) &=&-\frac 23D+\frac 13g_A^0  \label{a55} \\
G_A^s\left( 0\right) &=&-\frac 13\left( 3F-D\right) +\frac 13g_A^0  \nonumber
\end{eqnarray}
The simple picture of a proton having only non-strange valence $u$ and $d$
quarks require that 
\begin{eqnarray}
F_2^s\left( 0\right) &=&0\,\,\,\Longrightarrow \,\,\,\kappa _0=\kappa
_p+\kappa _n  \nonumber \\
G_A^s\left( 0\right) &=&0\,\,\,\Longrightarrow \,\,\,g_A^0=3F-D  \label{a56}
\end{eqnarray}
Thus strangeness in proton means that the value of $\kappa _0$ and $g_A^0$
are different from those given in Eq. (\ref{a56}). To answer this question
experimentally, the parity violating elastic scattering of polarized
(unpolarized) electrons on unpolarized (polarized) protons are of
importance. The interference between the photon exchange and $Z$-exchange
result in the form factors $F_2^Z$ and $G_A^Z$ which depend on the singlet
anomalous magnetic moment $\kappa _0$ and the isosinglet axial vector
constant $g_A^0$ [cf. Eqs. (\ref{51}) and (\ref{55})]. It may be noted that
electromagnetic form factors $F_2^{\gamma p}$ is independent of $\kappa _0$.
The experimental results both in polarized deep inelastic lepton--nucleon
scattering and for elastic neutrino proton scattering are consistent with $%
g_A^0=0$. In fact, the latter experiments \cite{PRD35-785} gives $%
g_A^0=0.12\pm 0.23$. This trend that singlet form factors is zero is
consistent with our discussion in section III. $g_A^0=0$ and $\kappa _0=0$,
implies [Eqs. (\ref{a55}) and (\ref{50})] 
\begin{eqnarray}
G_A^s\left( 0\right) &=&-\left( G_A^u\left( 0\right) +G_A^d\left( 0\right)
\right)  \nonumber \\
F_2^s\left( 0\right) &=&-\left( F_2^u\left( 0\right) +F_2^d\left( 0\right)
\right) =-\left( \kappa _p+\kappa _n\right)  \label{a57}
\end{eqnarray}

\section{Parity-violating unpolarized scattering on polarized protons}

In parity violating electron scattering, the weak and electromagnetic
interactions interfere. The unpolarized electron scattering on polarized
nucleons can be described in terms of six spin-dependent structure functions
(if we neglect the contribution to the scattering cross-section which goes
to zero as $m_e\rightarrow 0$). These structure function \cite{PRD38-3390}
are given by 
\begin{mathletters}
\begin{eqnarray}
\tilde{S}_{\mu \nu } &=&\left( 2\pi \right) ^3\frac{p_0}m\int
d^4z\,\,e^{-iq\cdot z}\left\langle p,n\left| \left[ J_\mu ^{\text{em}}\left(
z\right) ,J_\nu ^Z\left( 0\right) \right] \right| p,n\right\rangle  \nonumber
\\
&=&\tilde{S}_{\mu \nu }^{\left( -\right) }+\tilde{S}_{\mu \nu }^{\left(
+\right) }  \label{a58}
\end{eqnarray}
where 
\begin{eqnarray}
\tilde{S}_{\mu \nu }^{\left( -\right) } &=&-\frac{2\pi }m\tilde{G}%
_1^e\epsilon _{\mu \nu \rho \sigma }q_\rho n_\sigma +\frac{2\pi }{m^3}\tilde{%
G}_2^e\epsilon _{\mu \nu \rho \sigma }q_\rho \left( p\cdot qn_\sigma -n\cdot
qp_\sigma \right) -\frac{2\pi }m\tilde{G}_3^e\epsilon _{\mu \nu \rho \sigma
}p_\rho n_\sigma  \label{58a} \\
\tilde{S}_{\mu \nu }^{\left( +\right) } &=&-\frac{2\pi }m\tilde{H}_2^e\left(
p_\mu n_\nu +p_\nu n_\mu \right) -\frac{2\pi }m\tilde{H}_3^e\delta _{\mu \nu
}n\cdot q-\frac{2\pi }{m^3}\tilde{H}_4^ep_\mu p_\nu n\cdot q  \label{58b}
\end{eqnarray}
$n$ is the polarization vector of nucleon $\left( n^2=n_\mu n_\mu =1\text{, }%
p_\mu n_\mu =p\cdot n=0\right) $. Since we are considering unpolarized
electron, the lepton part $\tilde{L}_{\mu \nu }$ is given by (cf. Eq. (\ref
{43})) 
\end{mathletters}
\begin{eqnarray}
\tilde{L}_{\mu \nu } &=&\frac{v_e}{8m_e^2}\left[ k_\mu ^{\prime }k_\nu
+k_\nu ^{\prime }k_\mu -\delta _{\mu \nu }k\cdot k^{\prime }\right] +\frac{%
a_e}{8m_e^2}\epsilon _{\mu \nu \alpha \beta }k_\alpha k_\beta ^{\prime } 
\nonumber \\
&=&\tilde{L}_{\mu \nu }^{\left( +\right) }+\tilde{L}_{\mu \nu }^{\left(
-\right) }  \label{59}
\end{eqnarray}
The scattering cross-section is given by 
\begin{equation}
\frac{d^2\sigma }{dq^2d\nu }=\frac{4\pi \alpha }{q^2}\frac{8G_F}{\sqrt{2}E^2}%
\frac{m_e^2}{8\pi ^2}\left[ \tilde{L}_{\nu \mu }^{\left( -\right) }\tilde{S}%
_{\mu \nu }^{\left( -\right) }+\tilde{L}_{\nu \mu }^{\left( +\right) }\tilde{%
S}_{\mu \nu }^{\left( +\right) }\right]  \label{60}
\end{equation}
>From Eqs. (55) and (\ref{59}), we get 
\begin{eqnarray}
\tilde{L}_{\nu \mu }^{\left( -\right) }\tilde{S}_{\mu \nu }^{\left( -\right)
} &=&\frac{a_e}{8m_e^2}\frac{4\pi }m\left\{ \tilde{G}_1^e\left( k\cdot
qk^{\prime }\cdot n-k^{\prime }\cdot qk\cdot n\right) \right.  \nonumber \\
&&-\frac{\tilde{G}_2^e}{m^2}\left[ p\cdot q\left( k\cdot qk^{\prime }\cdot
n-k^{\prime }\cdot qk\cdot n\right) +q\cdot n\left( k\cdot pk^{\prime }\cdot
q-k\cdot qk^{\prime }\cdot p\right) \right]  \nonumber \\
&&\left. +\tilde{G}_3^e\left( k\cdot pk^{\prime }\cdot n-k^{\prime }\cdot
pk\cdot n\right) \right\}  \label{61} \\
\tilde{L}_{\nu \mu }^{\left( +\right) }\tilde{S}_{\mu \nu }^{\left( +\right)
} &=&-\frac{v_e}{8m_e^2}\frac{4\pi }m\left\{ \tilde{H}_2^e\left( p\cdot
k^{\prime }k\cdot n+p\cdot kk^{\prime }\cdot n\right) -\tilde{H}_3^ek\cdot
k^{\prime }q\cdot n 
\begin{tabular}{c}
$\,\,\,\,\,\,\,$ \\ 
$\,\,\,\,\,\,\,\,$%
\end{tabular}
\right.  \nonumber \\
&&\left. +\frac{\tilde{H}_4^e}{m^2}\left( p\cdot kp\cdot k^{\prime }-\frac 12%
p^2k\cdot k^{\prime }\right) q\cdot n\right\}  \label{62}
\end{eqnarray}
In the laboratory frame, it is convenient to take $\vec{k}$ along z--axis.
Thus in the Lab frame (for $E\gg m_e$): 
\begin{eqnarray}
k &\equiv &\left( 0,0,E,iE\right) ,  \nonumber \\
k^{\prime } &=&E^{\prime }\left( \sin \theta \cos \phi ,\sin \theta \sin
\phi ,\cos \theta ,i\right) ,  \nonumber \\
p &\equiv &\left( 0,im\right)  \label{63}
\end{eqnarray}
We will confine ourself to longitudnal polarization of proton: 
\begin{equation}
n=\left( 0,0,\lambda ,0\right)
;\,\,\,\,\,\,\,\,\,\,\,\,\,\,\,\,\,\,\,\lambda =\pm 1  \label{64}
\end{equation}

Then from Eq. (\ref{60}), using Eqs. (\ref{61}), (\ref{62}) and (\ref{63}),
we get \cite{PRD38-3390} 
\begin{eqnarray}
\frac{d^2\sigma ^{\overrightarrow{\leftarrow }}}{dq^2d\nu } &=&\lambda \frac{%
\alpha G_F}{\sqrt{2}q^2}\frac 1{mE^2}\left\{ v_e\left[ 2m\tilde{H}%
_2^eEE^{\prime }\left( 1+\cos \theta \right) -\tilde{H}_3^eq^2\left(
E-E^{\prime }\cos \theta \right) \right. \right.  \nonumber \\
&&-\left. \tilde{H}_4^e\left( 2EE^{\prime }-\frac{q^2}2\right) \left(
E-E^{\prime }\cos \theta \right) \right]  \nonumber \\
&&+\left. a_eq^2\left[ \tilde{G}_1^e\left( E+E^{\prime }\cos \theta \right) -%
\tilde{G}_2^e\frac{q^2}m+m\tilde{G}_3^e\right] \right\}  \label{65}
\end{eqnarray}
For elastic scattering the structure functions are given by\footnote{%
There were some errors in this paper; these have been corrected here} 
\cite{PRD42-794} 
\begin{eqnarray}
\tilde{H}_2^e &=&G_A^Z\left( F_1^\gamma -\tau F_2^\gamma \right) \delta
\left( \nu -\frac{q^2}{2m}\right) ,  \nonumber \\
\tilde{H}_3^e &=&-G_A^Z\left( F_1^\gamma +F_2^\gamma \right) \delta \left(
\nu -\frac{q^2}{2m}\right) ,  \nonumber \\
\tilde{H}_4^e &=&-G_A^ZF_2^\gamma \delta \left( \nu -\frac{q^2}{2m}\right)
\label{66} \\
\tilde{G}_1^e &=&\frac 12\left[ F_1^\gamma \left( F_1^Z+F_2^Z\right)
+F_1^Z\left( F_1^\gamma +F_2^\gamma \right) \right] \delta \left( \nu -\frac{%
q^2}{2m}\right)  \nonumber \\
\tilde{G}_2^e &=&-\frac 14\left[ F_2^\gamma \left( F_1^Z+F_2^Z\right)
+F_2^Z\left( F_1^\gamma +F_2^\gamma \right) \right] \delta \left( \nu -\frac{%
q^2}{2m}\right)  \label{67}
\end{eqnarray}
Using Eqs. (\ref{66}) and (\ref{67}), we get the scattering cross-section 
\cite{PRD42-794,9811055} 
\begin{eqnarray}
\frac{d\sigma ^{\overrightarrow{\leftarrow }}}{dq^2} &=&\lambda \frac{G_Fq^2%
}{\sqrt{2}\pi \alpha }\left( \frac{d\sigma }{dq^2}\right) _{\text{Mott}%
}\left\{ v_eG_A^Z\left[ \left( F_1^\gamma +\tau \frac mEF_2^\gamma \right)
+2\tau \left( 1+\frac mE\right) \tan ^2\frac \theta 2\left( F_1^\gamma
+F_2^\gamma \right) \right] \right.  \nonumber \\
&&\left. +2a_e\tan ^2\frac \theta 2\left[ \frac 12\left( \frac Em-\tau \frac 
mE\right) \left( F_1^\gamma \left( F_1^Z+F_2^Z\right) +F_1^Z\left(
F_1^\gamma +F_2^\gamma \right) \right) -\tau \left( F_1^\gamma
F_1^Z-F_2^ZF_2^\gamma \right) \right] \right\}  \nonumber \\
&&  \label{68}
\end{eqnarray}
Now using Eqs. (\ref{13}) and (\ref{14}), the longitudnal asymmetry for the
scattering of unpolarized electrons on polarized protons can be written 
\begin{eqnarray}
{\cal A}_L^p &=&\frac{d\sigma ^{\leftarrow }-d\sigma ^{\rightarrow }}{%
d\sigma ^{\leftarrow }+d\sigma ^{\rightarrow }}  \nonumber \\
&=&-\frac{G_Fq^2}{\sqrt{2}\pi \alpha }\frac 1{\Sigma _\gamma }\left\{
v_eG_A^Z\left[ \epsilon \left( G_E\left( 1-\tau \frac mE\right) +\tau
G_M\left( 1+\frac mE\right) \right) +\left( 1-\epsilon \right) \left( 1+%
\frac mE\right) G_M\right] \right.  \nonumber \\
&&+\frac{a_e}2\frac{1-\epsilon }{1+\tau }\left[ \left( \frac Em-\tau \frac mE%
\right) \left( G_EG_M^Z+G_E^ZG_M+2\tau G_MG_M^Z\right) \right.  \nonumber \\
&&\left. \left. -2\tau \left( G_EG_M^Z+G_E^ZG_M-\left( 1-\tau \right)
G_MG_M^Z\right) \right] \right\}  \label{69}
\end{eqnarray}
where we have used the Sachs form factors define in Eq. (\ref{08}) and 
\begin{equation}
\Sigma _\gamma =\tau G_M^2+\epsilon G_E^2  \label{70}
\end{equation}
Recently it has been pointed out that internal structure of the proton can
be investigated with polarization transfer. For one photon exchange, the
scattering of longitunally polarized electrons results in a transfer of
polarization to the recoil proton (see Sec. VIII)

Let us now discuss the polarization of recoil proton for our case. For this
case, the four vector $n_\mu $ refers to the recoil proton, i.e. 
\begin{eqnarray}
n\cdot p^{\prime } &=&0,  \nonumber \\
p^{\prime } &=&k-k^{\prime }+p,  \nonumber \\
E_{p^{\prime }} &=&E-E^{\prime }+m  \label{e71} \\
{\bf p}^{\prime } &=&{\bf k}-{\bf k}^{\prime }  \nonumber \\
&\equiv &\left( -E^{\prime }\cos \phi \sin \theta ,-E^{\prime }\sin \phi
\sin \theta ,E-E^{\prime }\cos \theta \right)  \nonumber \\
&=&\left| {\bf p}^{\prime }\right| \left( \cos \phi \sin \beta ,\sin \phi
\sin \beta ,\cos \beta \right)
\end{eqnarray}
Hence we have 
\begin{eqnarray}
\left| {\bf p}^{\prime }\right| &=&2m\sqrt{\tau \left( 1+\tau \right) } 
\nonumber \\
E_{p^{\prime }} &=&m\left( 1+2\tau \right)  \label{73} \\
-E^{\prime }\sin \theta &=&\left| {\bf p}^{\prime }\right| \sin \beta =2m%
\sqrt{\tau \left( 1+\tau \right) }\sin \beta  \nonumber \\
E-E^{\prime }\cos \theta &=&m\tau \left( 1+\frac mE\right) =2m\sqrt{\tau
\left( 1+\tau \right) }\cos \beta  \label{74}
\end{eqnarray}
For transverse polarization of recoil proton, 
\begin{eqnarray}
{\bf p}^{\prime }\cdot {\bf n} &=&0={\bf q}\cdot {\bf n},\,\,\,n_0=0 \\
{\bf n} &\equiv &\left( \cos \phi \cos \beta ,\sin \phi \sin \beta ,-\sin
\beta \right)  \label{75}
\end{eqnarray}
Then from Eqs. (\ref{60}--\ref{62}, \ref{66}, \ref{67}), using Eqs. (\ref{73}%
,\ref{74},\ref{75}), we get 
\begin{eqnarray}
\left( \frac{d\sigma }{dq^2}\right) _{\text{T recoil}} &=&\frac{G_F}{\sqrt{2}%
}\frac{q^2}{\pi \alpha }\left( \frac{d\sigma }{dq^2}\right) _{\text{Mott}}%
\frac{\tan \left( \theta /2\right) }{\sqrt{\tau \left( 1+\tau \right) }} 
\nonumber \\
&&\times \left\{ v_e\left( \frac Em\right) \left( 1-\tau \frac mE\right)
G_A^ZG_E+a_e\tau \left( G_EG_M^Z+G_{_E}^ZG_M\right) \right\}  \label{76}
\end{eqnarray}
For longitudnal polarization of recoil proton, 
\begin{eqnarray}
{\bf n} &\equiv &-\left( \cos \phi \sin \beta ,\sin \phi \sin \beta ,\cos
\beta \right)  \nonumber \\
n_0 &=&\frac{{\bf p}^{\prime }\cdot {\bf n}}{E_{p^{\prime }}}  \label{77}
\end{eqnarray}
Again, from Eqs. (\ref{60}--\ref{62}, \ref{66}, \ref{67}), using Eqs. (\ref
{73}--\ref{75}), and retaining only the contribution of ${\bf n}$, we get 
\begin{eqnarray}
\left( \frac{d\sigma }{dq^2}\right) _{\text{L recoil}} &=&-\frac{G_F}{\sqrt{2%
}}\frac{q^2}{\pi \alpha }\left( \frac{d\sigma }{dq^2}\right) _{\text{Mott}}%
\frac 1{\epsilon \left( 1+\tau \right) }\sqrt{\frac \tau {1+\tau }} 
\nonumber \\
&&\times \left\{ v_eG_A^Z\left[ \frac \epsilon {1+\tau }\left( G_M+\tau
G_E\right) -\left( 1-\epsilon \right) G_M\right] +a_e\left( 1-\epsilon
\right) \left( \frac{E+E^{\prime }}{2m}\right) G_MG_M^Z\right\}  \label{78}
\end{eqnarray}
Hence for the longitudnal polarization of the recoil proton 
\begin{eqnarray}
I_L &\equiv &\left( \frac{d\sigma }{dq^2}\right) _{\text{L recoil}}/\left( 
\frac{d\sigma }{dq^2}\right) _{\text{em}}  \nonumber \\
&=&-\frac{G_F}{\sqrt{2}}\frac{q^2}{\pi \alpha }\frac 1{\tau G_M^2+\epsilon
G_E^2}\sqrt{\frac \tau {1+\tau }}  \nonumber \\
&&\times \left\{ v_eG_A^Z\left[ \frac \epsilon {1+\tau }\left( G_M+\tau
G_E\right) -\left( 1-\epsilon \right) G_M\right] +a_e\left( 1-\epsilon
\right) \left( \frac Em-\tau \right) G_MG_M^Z\right\}  \label{79}
\end{eqnarray}
It may be noted that electroweak interference due to one-photon exchange and
Z-exchange can induce polarization in the recoil proton, although the lepton
beam is unpolarized. The V--A term in the lepton sector i.e. the term $%
\epsilon _{\mu \nu \alpha \beta }k_\alpha k_\beta ^{\prime }$ which is
similar in content to $\epsilon _{\mu \nu \alpha \beta }q_\alpha s_\beta $ ($%
s_\mu $, polarization of lepton) contracted with the antisymmetric part of
hadronic sector [cf. Eq. (\ref{58a})] gives rise to the term with
coefficient $a_e$ in Eq. (\ref{76}) and (\ref{78}). The first term with $v_e$
coefficient in Eqs. (\ref{76}) and (\ref{78}) arises due to the contraction
of symmetric part of leptonic tensor with the symmetric part of hadronic
tensor in Eq. (\ref{58b}). The symmetric part of hadronic tensor arises due
to V--A term in hadronic sectror which would vanish in the absence of
electro-weak interference. Experimentally it may be possible to detect the
polarization of recoil proton in the unpolarized electron proton scattering.
The second term in Eq. (\ref{79}) is similar in character to that of
polarization transfer from lepton to proton (Sec. VIII).

\section{Parity--Violating polarized electrons scattering on unpolarized
protons}

For this case, the relevant structure functions are given by 
\begin{eqnarray}
\tilde{J}_{\mu \nu } &=&\left( 2\pi \right) ^3\frac{p_0}m\int e^{-iq\cdot
z}\left\langle p\left| \left[ J_\mu ^{\text{em}}\left( z\right) ,J_\nu
^Z\left( 0\right) \right] \right| p\right\rangle  \nonumber \\
&=&2\pi \left\{ \frac{\tilde{W}_2}{m^2}\left[ p_\mu p_\nu -\frac{p\cdot q}{%
q^2}\left( p_\mu q_\nu +p_\nu q_\mu \right) +\frac{\left( p\cdot q\right) ^2%
}{q^4}q_\mu q_\nu \right] \right.  \nonumber \\
&&\left. +\left( \delta _{\mu \nu }-\frac{q_\mu q_\nu }{q^2}\right) \tilde{W}%
_1+\frac 1{2m^2}\epsilon _{\mu \nu \rho \sigma }p_\rho q_\sigma \tilde{W}%
_3\right]  \label{80}
\end{eqnarray}
For polarized electrons, the lepton part containing the electron
polarization part is given by 
\begin{equation}
\left( \tilde{L}_{\mu \nu }\right) _s=-v_e\frac 1{4m_e}\epsilon _{\mu \nu
\alpha \beta }q_\alpha s_\beta -\frac{a_e}{4m_e}\left[ k_\mu ^{\prime }s_\nu
+k_\nu ^{\prime }s_\mu -\delta _{\mu \nu }k^{\prime }\cdot s+k^{\prime
}\rightarrow k\right]  \label{81}
\end{equation}
where $s_\mu $ is the polarization vector of electron. The scattering
cross-section containing only polarized part is given by 
\begin{eqnarray}
\frac{d^2\sigma }{dq^2d\nu } &=&\frac{4\pi \alpha }{q^2}\frac{8G_F}{\sqrt{2}%
E^2}\frac{m_e^2}{8\pi ^2}\tilde{L}_{\nu \mu }\tilde{J}_{\mu \nu }  \nonumber
\\
&=&-\frac \alpha {q^2}\frac{G_F}{\sqrt{2}}\frac{m_e}{m^2}\left\{ v_e\left(
q^2p\cdot s+p\cdot qk^{\prime }\cdot s\right) \frac{\tilde{W}_3}{m^2}\right.
\nonumber \\
&&\left. +a_e\left[ \frac{\tilde{W}_2}{m^2}\left( 2p\cdot k^{\prime }p\cdot
n+m^2k^{\prime }\cdot n\right) -2\tilde{W}_1k^{\prime }\cdot n\right]
\right\}  \label{82}
\end{eqnarray}
we note 
\begin{eqnarray}
k\cdot s &=&0  \nonumber \\
k &=&\left( 0,0,1,i\right)  \label{83}
\end{eqnarray}
For longitudnally polarized electron beam 
\[
s\equiv \lambda \left( 0,0,\frac E{m_e},\frac{iE}{m_e}\right) 
\]
Hence for this case 
\begin{eqnarray}
\frac{d^2\sigma }{dq^2d\nu } &=&\lambda \frac \alpha {q^2}\frac{G_F}{\sqrt{2}%
}\frac 1{E^2}\left\{ v_e\tilde{W}_3\frac{E+E^{\prime }}{2m}q^2-a_e\left[ 
\tilde{W}_2EE^{\prime }\left( 1+\cos \theta \right) +\tilde{W}_1EE^{\prime
}\left( 1-\cos \theta \right) \right] \right\}  \label{85} \\
&=&\frac \lambda {2\pi }\frac{G_F}{\sqrt{2}}\frac{q^2}\alpha \left( \frac{%
d\sigma }{dq^2}\right) _{\text{Mott}}\left\{ v_e\tilde{W}_3\frac{E+E^{\prime
}}m\tan ^2\frac \theta 2-a_e\left[ \tilde{W}_2+2\tilde{W}_1\tan ^2\frac 
\theta 2\right] \right\}  \label{86}
\end{eqnarray}
For elastic scattering 
\begin{eqnarray}
\tilde{W}_1 &=&2\tau \left( F_1^\gamma +F_2^\gamma \right) \left(
F_1^Z+F_2^Z\right) \delta \left( \nu -\frac{q^2}{2m}\right)  \nonumber \\
\tilde{W}_2 &=&2\left( F_1^\gamma F_1^Z+\tau F_2^\gamma F_2^Z\right) \delta
\left( \nu -\frac{q^2}{2m}\right)  \nonumber \\
\tilde{W}_3 &=&-2G_A^Z\left( F_1^\gamma +F_2^\gamma \right)  \label{87}
\end{eqnarray}
Hence the elastic scattering cross-section for longitudnally polarized
electrons on unpolarized protons is given by \cite{PLB219-140} 
\begin{eqnarray}
\frac{d\sigma ^{\overrightarrow{\leftarrow }}}{dq^2} &=&\pm \frac{G_F}{\sqrt{%
2}}\frac{q^2}{\pi \alpha }\left( \frac{d\sigma }{dq^2}\right) _{\text{Mott}%
}\left\{ v_e\left[ -2G_A^Z\left( F_1^\gamma +F_2^\gamma \right) \frac{%
E-m\tau }m\tan ^2\frac \theta 2\right] \right.  \nonumber \\
&&-a_e\left[ \left( F_1^\gamma F_1^Z+\tau F_2^\gamma F_2^Z\right) +2\tau
\tan ^2\frac \theta 2\left( F_1^\gamma +F_2^\gamma \right) \left(
F_1^Z+F_2^Z\right) \right]  \label{88}
\end{eqnarray}
For this case the asymmetry is given by 
\begin{eqnarray}
{\cal A}^e &=&\frac{d\sigma ^{\rightarrow }-d\sigma ^{\leftarrow }}{d\sigma
^{\rightarrow }+d\sigma ^{\leftarrow }}  \nonumber \\
&=&-\frac{G_F}{\sqrt{2}}\frac{q^2}{\pi \alpha }\frac 1{\Sigma _\gamma }%
\left\{ v_e\left( 1-\epsilon \right) \left( \frac Em-\tau \right)
G_A^ZG_M\right.  \nonumber \\
&&\left. +a_e\left[ \epsilon \left( G_E^ZG_E+\tau G_M^ZG_M\right) +\tau
\left( 1-\epsilon \right) G_M^ZG_M\right] \right\}  \label{89}
\end{eqnarray}
It may be noted that asymmetry ${\cal A}^e$ given in Eq. (\ref{89}) and
longitudnal asymmetry ${\cal A}_L^p$ given in Eq. (\ref{69}) complement each
other.

\section{Measurement of the elastic form factors $G_A^Z$, $G_M^Z$}

The elastic scattering of electrons on proton target involves the form
factors $G_E$, $G_M$, $G_E^Z$, $G_M^Z$ and $G_A^Z$. These form factors
characterize the internal structure of proton. In particular $G_A^Z$ and $%
G_M^Z$ contain both octet and singlet part. Thus the measurent of $G_A^Z$
and $G_M^Z$ would give the information about the singlet form factors which
in turn determines the strange content of the proton. The electromagnetic
form factors follow the dipole form 
\begin{equation}
G_E=\frac 1{\mu _p}G_M=G_D=\frac 1{\left[ 1+q^2/0.71\text{ GeV}^2\right] ^2}
\label{e90}
\end{equation}
We take the dipole form for $G_A^Z$, $G_E^Z$ and $G_M^Z$: 
\begin{eqnarray}
G_E^Z &=&\frac{G_E^Z\left( 0\right) }{\left[ 1+q^2/m_V^2\right] ^2} 
\nonumber \\
G_M^Z &=&\frac{G_M^Z\left( 0\right) }{\left[ 1+q^2/m_V^2\right] ^2} 
\nonumber \\
G_A^Z &=&\frac{G_A^Z\left( 0\right) }{\left[ 1+q^2/m_A^2\right] ^2}
\label{e91}
\end{eqnarray}
We will assume that 
\begin{eqnarray}
m_V &\equiv &m_V^3=m_V^8=m_V^0=0.84\text{ GeV, }m_V^2=0.71\text{ GeV}^2 
\nonumber \\
m_A &\equiv &m_A^3=m_A^8=m_A^0=1.03\text{ GeV, }m_A^2=1.06\text{ GeV}^2
\label{e92}
\end{eqnarray}
The equality of $m_V^3=m_V^8$ follows from the electromagnetic form factors;
similarly $m_A^3=m_A^8$ follows from $\beta $-decay. The particular value
for $m_A$ is taken from Ref. \cite{PRD35-785}. Now 
\begin{equation}
G_A^Z\left( 0\right) =\frac 12\left[ \left( F+D\right) +\left( F-\frac 13%
D\right) -\frac 13g_A^0\right] ,  \label{e93}
\end{equation}
where experimentally 
\begin{eqnarray}
F+D &=&1.2670\pm 0.0035  \nonumber \\
F-\frac 13D &=&0.25\pm 0.05  \label{e94}
\end{eqnarray}
EMC and elastic neutrino scattering data are consistent with $g_A^0\approx 0$%
. For $g_A^0=0$, 
\[
G_A^s\left( 0\right) =-\left( G_A^u\left( 0\right) +G_A^d\left( 0\right)
\right) 
\]
where as $g_A^0=3F-D$, gives 
\begin{equation}
G_A^s\left( 0\right) =0  \label{e95}
\end{equation}
For the vector form factors 
\begin{eqnarray}
G_E^Z\left( 0\right) &=&\frac 12\left( 1-4x_W\right) =0.036;\,\,\,\,\,\,\,%
\text{for }x_W=0.2322  \nonumber \\
G_M^Z\left( 0\right) &=&\left( 1-2x_W\right) \mu _p-\frac 12\left( 1+\kappa
_0\right)  \label{e96}
\end{eqnarray}
Again the singlet anamolous magnetic moment $\kappa _0$ is unknown. For $%
\kappa _0=0$: 
\[
G_M^s\left( 0\right) =-\left( G_M^u\left( 0\right) +G_M^D\left( 0\right)
\right) 
\]
where as for $\kappa _0=\kappa _p+\kappa _n$%
\begin{equation}
G_M^s\left( 0\right) =0  \label{e97}
\end{equation}
In our analysis of parity violating scattering, we will use two sets of
values

\begin{itemize}
\item[set (i)]  . 
\begin{eqnarray}
G_A^Z\left( 0\right) &=&\frac 12\left[ \left( F+D\right) +\left( F-\frac 13%
D\right) \right] =0.76  \nonumber \\
G_M^Z\left( 0\right) &=&\left( 1-2x_W\right) \mu _p-\frac 12=0.99
\label{e98}
\end{eqnarray}

\item[set (ii)]  . 
\begin{eqnarray}
G_A^Z\left( 0\right) &=&\frac 12\left[ \left( F+D\right) +\left( F-\frac 13%
D\right) -\left( F-\frac 13D\right) \right] =0.63  \nonumber \\
G_M^Z\left( 0\right) &=&\left( 1-2x_W\right) \mu _p-\frac 12\left( 1+\kappa
_p+\kappa _n\right) =1.05  \label{e99}
\end{eqnarray}
\end{itemize}

We note that since $G_E^Z\left( 0\right) /G_M^Z\left( 0\right) \approx 0.036$%
, it is a good approximation to neglect the contribution of $G_E^Z$ as
compared with $G_M^Z$.

Using Rosenbulth method, it is possible to extract the form factors by
measuring the longitudnal proton and electron asymmetries as well as recoil
proton polarizations at a fixed $q^2$ over the range of $\epsilon $ values
that are obtained by changing the beam energy and scattering angle. In order
to use this technique, we re-express the asymmetries and recoil polarization
given in Eqs. (\ref{69}), (\ref{89}), (\ref{79}) and (\ref{78}) in terms of $%
\epsilon $ and $\tau $. For this purpose we note 
\begin{eqnarray}
\tan ^2\frac \theta 2 &=&\frac{1-\epsilon }{2\epsilon \left( 1+\tau \right) }
\label{e100} \\
\epsilon &=&\frac{E^2-2mE\tau -m^2\tau }{E^2-2mE\tau +m^2\tau +2m^2\tau ^2}
\label{e101} \\
\frac Em-\tau &=&\sqrt{\tau \left( 1-\tau \right) }\sqrt{\frac{1+\epsilon }{%
1-\epsilon }}  \label{e102}
\end{eqnarray}
Hence from Eqs. (\ref{69}), (\ref{89}), (\ref{79}) and (\ref{78}) we get 
\begin{eqnarray}
&&\left. {\cal A}_L^p=-\left( \frac{G_Fq^2}{\sqrt{2}\pi \alpha }\right)
\left( \frac 1{\tau G_M^2+\epsilon G_E^2}\right) \left( \frac 1{1+\epsilon
+2\epsilon \tau }\right) \right.  \nonumber \\
&&\times \left\{ v_eG_A^Z\left[ 2\epsilon ^2\left( 1+\tau \right) G_E+2\tau
G_M\left( 1-\epsilon +\epsilon \tau \right) +\sqrt{\tau \left( 1+\tau
\right) \left( 1-\epsilon ^2\right) }\left( -\epsilon G_E+\left( 1-\epsilon
+\epsilon \tau \right) G_M\right) \right] \right.  \nonumber \\
&&+a_e\left[ \frac{1-\epsilon }{1+\tau }\left( \left(
G_EG_M^Z+G_E^ZG_M\right) \left( \epsilon -\tau -2\epsilon \tau ^2\right)
+\tau G_MG_M^Z\left( 2\epsilon \left( 1+\tau \right) +\left( 1-\tau \right)
\left( 1+\epsilon +2\epsilon \tau \right) \right) \right) \right.  \nonumber
\\
&&\left. \left. +\epsilon \sqrt{\tau \left( 1+\tau \right) \left( 1-\epsilon
^2\right) }\left( G_EG_M^Z+G_E^ZG_M+2\tau G_MG_M^Z\right) \right] \right\}
\label{e103} \\
&&\left. {\cal A}^e=-\left( \frac{G_Fq^2}{\sqrt{2}\pi \alpha }\right) \left( 
\frac 1{\tau G_M^2+\epsilon G_E^2}\right) \right.  \nonumber \\
&&\times \left\{ v_e\sqrt{\tau \left( 1+\tau \right) \left( 1-\epsilon
^2\right) }G_A^ZG_M+a_e\left[ \epsilon G_E^ZG_E+\tau G_M^ZG_M\right] \right\}
\label{e104} \\
&&\left. {\cal I}_L=-\left( \frac{G_Fq^2}{\sqrt{2}\pi \alpha }\right) \left( 
\frac 1{\tau G_M^2+\epsilon G_E^2}\right) \sqrt{\frac \tau {1+\tau }}\right.
\nonumber \\
&&\times \left\{ v_eG_A^Z\left[ \frac \epsilon {1+\tau }\left( G_M+\tau
G_E\right) -\left( 1-\epsilon \right) G_M\right] +a_e\sqrt{\tau \left(
1+\tau \right) \left( 1-\epsilon ^2\right) }G_MG_M^Z\right\}  \label{e105} \\
&&\left. {\cal I}_T=-\left( \frac{G_Fq^2}{\sqrt{2}\pi \alpha }\right) \left( 
\frac 1{\tau G_M^2+\epsilon G_E^2}\right) \sqrt{\frac 1{2\tau }}\right. 
\nonumber \\
&&\times \left\{ v_eG_A^ZG_E\frac{2\epsilon \left( 1+\tau \right) }{%
1+\epsilon +2\epsilon \tau }\left[ \sqrt{\tau \left( 1+\tau \right) \epsilon
\left( 1+\epsilon \right) }+\sqrt{\epsilon \left( 1-\epsilon \right) }%
\right] \right.  \nonumber \\
&&\left. +a_e\tau \sqrt{\epsilon \left( 1-\epsilon \right) }\left(
G_EG_M^Z+G_E^ZG_M\right) \right\}  \label{e106}
\end{eqnarray}
First we note that since the form factor $G_A^Z$ is multiplied by $%
v_e=-\left( 1-4\sin ^2\theta _W\right) =-0.07$, its contribution is
suppressed compared to the terms multiplied by $a_e=-1$. Thus overall
contribution of $G_A^Z$ is suppressed except for $\epsilon =1$ in Eqs. (\ref
{e103}), (\ref{e105}) and (\ref{e106}). For $\epsilon =1$, we get from these
equations 
\begin{eqnarray}
{\cal A}_L^p\left( \epsilon =1\right) &=&-\left( \frac{G_Fq^2}{\sqrt{2}\pi
\alpha }\right) \left( \frac{G_M^2}{\tau G_M^2+G_E^2}\right) \frac 1{2\left(
1+\tau \right) }\left\{ v_e\frac{G_A^Z}{G_M}\left[ 2\left( 1+\tau \right) 
\frac{G_E}{G_M}+2\tau ^2\right] \right\}  \label{e107} \\
{\cal I}_L\left( \epsilon =1\right) &=&-\left( \frac{G_Fq^2}{\sqrt{2}\pi
\alpha }\right) \left( \frac{G_M^2}{\tau G_M^2+\epsilon G_E^2}\right) \sqrt{%
\frac \tau {1+\tau }}\left\{ v_e\frac{G_A^Z}{G_M}\left[ \left( 1+\tau \frac{%
G_E}{G_M}\right) \frac 1{1+\tau }\right] \right\}  \label{e108} \\
{\cal I}_T\left( \epsilon =1\right) &=&\frac{G_Fq^2}{\sqrt{2}\pi \alpha }%
\left( \frac{G_M^2}{\tau G_M^2+\epsilon G_E^2}\right) \left\{ v_e\frac{G_A^Z%
}{G_M}\frac{G_E}{G_M}\sqrt{1+\tau }\right\}  \label{e109}
\end{eqnarray}
On the other hand from Eq. (\ref{e104}), we get at $\epsilon =1$%
\begin{equation}
{\cal A}_e\left( \epsilon =1\right) =-\left( \frac{G_Fq^2}{\sqrt{2}\pi
\alpha }\right) \left( \frac{G_M^2}{\tau G_M^2+G_E^2}\right) a_e\frac{G_M^Z}{%
G_M}  \label{e110}
\end{equation}
It is thus clear from Eqs. (\ref{e107}--\ref{e110}), that it is possible to
extract $G_A^Z/G_M$ and $G_M^Z/G_M$ using Rosenbluth method. In any case we
have plotted ${\cal A}^p$, ${\cal A}_e$, ${\cal I}_L$, and ${\cal I}_T$,
versus $\epsilon $ for $q^2=0.5$ GeV, 1 GeV, 2.64 GeV, 3.2 GeV, 5.6 GeV, and
10 GeV. In these plots we have taken 
\begin{eqnarray}
\frac{G_A^Z}{G_M} &=&\frac{G_A^Z\left( 0\right) }{\mu _p}\frac{\left(
1+q^2/0.71\text{ GeV}^2\right) ^2}{\left( 1+q^2/1.06\text{ GeV}^2\right) ^2}
\label{e111} \\
\frac{G_M^Z}{G_M} &=&\frac{G_M^Z\left( 0\right) }{\mu _p},\,\,\,\,\,\,\,\,%
\frac{G_E}{G_M}=\frac 1{\mu _p}  \label{e112}
\end{eqnarray}
For $G_A^Z\left( 0\right) $ and $G_M^Z\left( 0\right) $, we have used two
sets of values given in Eqs. (\ref{e98}) and (\ref{e99}). These plots are
shown in Figures 2, 3, 4 and 5. These figures can be used to extract the
form factors in future experiments.

\section{Polarized electrons scattering on polarized protons}

For this case, we replace $\tilde{S}_{\mu \nu }\rightarrow S_{\mu \nu }$ and 
$\tilde{G}_{1,2}^e\rightarrow \left( 1/2\right) G_{1,2}$ in Eq. (\ref{58b});
the other structure functions are not relevant for this case. Then the
scattering cross-section for polarized electrons on polarized protons is
given by 
\begin{equation}
\frac{d^2\sigma }{dq^2dv}=\frac 1\pi \left( \frac{d\sigma }{dq^2}\right) _{%
\text{Mott}}\frac{4m_e^2}{q^2}\tan ^2\frac \theta 2\left[ \left( L_{\nu \mu
}\right) _sS_{\mu \nu }\right] ,  \label{e113}
\end{equation}
where [cf. Eq. (\ref{82})] 
\begin{equation}
\left( L_{\nu \mu }\right) _s=-\frac 1{m_e}\epsilon _{\nu \mu \alpha \beta
}q_\alpha s_\beta  \label{e114}
\end{equation}
Thus 
\begin{equation}
\left( L_{\nu \mu }\right) _sS_{\mu \nu }=\frac 12\frac{2\pi }{mm_e}\left[
-G_1\left( q^2n\cdot s-n\cdot qs\cdot q\right) +\frac{G_2}{m^2}\left(
q^2p\cdot qn\cdot s-q^2n\cdot qp\cdot s\right) \right]  \label{e115}
\end{equation}
Hence for longitudnally polarized electrons, the scattering cross-section is
given by 
\begin{eqnarray}
\frac{d^2\sigma ^{\rightarrow }}{dq^2d\nu } &=&\left( \frac{d\sigma }{dq^2}%
\right) _{\text{Mott}}\frac{4E}{mq^2}\tan ^2\frac \theta 2  \nonumber \\
&&\times \left\{ -G_1\left( q^2\left( n_z-n_0\right) -n\cdot qE^{\prime
}\left( 1-\cos \theta \right) \right) +\frac{G_2}m\left( -q^2\nu \left(
n_z-n_0\right) +q^2n\cdot q\right) \right\}  \label{e116}
\end{eqnarray}
For longitudnally polarized protons, $\vec{n}=\lambda _n\vec{e}_z$, $n_0=0$, 
$\lambda _n=\pm 1$, we get 
\begin{equation}
\frac{d^2\sigma ^{\overrightarrow{\leftarrow }}}{dq^2d\nu }=-2\left( \frac{%
d\sigma }{dq^2}\right) _{\text{Mott}}\frac 1m\tan ^2\frac \theta 2\left[
G_1\left( E+E^{\prime }\cos \theta \right) -\frac{q^2}mG_2\right]
\label{e117}
\end{equation}
Now longitudnal spin--spin asymmetryis defined as: 
\begin{eqnarray}
{\cal A}_{\parallel } &=&\frac{d\sigma ^{\overrightarrow{\leftarrow }%
}-d\sigma ^{\overrightarrow{\rightarrow }}}{d\sigma ^{\overrightarrow{%
\leftarrow }}+d\sigma ^{\overrightarrow{\rightarrow }}}  \nonumber \\
&=&2\left( \frac{d\sigma }{dq^2}\right) _{\text{Mott}}\frac 1m\tan ^2\frac 
\theta 2\left[ G_1\left( E+E^{\prime }\cos \theta \right) -\frac{q^2}m%
G_2\right] /\frac{d^2\sigma }{dq^2d\nu }  \label{e118}
\end{eqnarray}
where [cf. Eq. (\ref{04})] 
\begin{eqnarray}
\frac{d^2\sigma }{dq^2d\nu } &=&\left( \frac{d\sigma }{dq^2}\right) _{\text{%
Mott}}W_1\left[ \frac{W_2}{W_1}+2\tan ^2\frac \theta 2\right]  \nonumber \\
&=&\left( \frac{d\sigma }{dq^2}\right) _{\text{Mott}}\frac{W_1}{\epsilon
\left( 1+\nu ^2/q^2\right) }\left[ 1+\epsilon R\right]  \nonumber \\
&=&\left( \frac{d\sigma }{dq^2}\right) _{\text{Mott}}\frac{2\tan ^2\frac 
\theta 2}{1-\epsilon }W_1\left[ 1+\epsilon R\right]  \label{e119}
\end{eqnarray}
where we have used 
\begin{eqnarray}
\left( 1+\frac{\nu ^2}{q^2}\right) \frac{W_2}{W_1}-1 &=&\frac{\sigma _L}{%
\sigma _T}=R  \label{e120} \\
\epsilon &=&\frac 1{1+2\left( 1+\nu ^2/q^2\right) \tan ^2\frac \theta 2}
\label{e121}
\end{eqnarray}
Hence longitudnal spin--spin asymmetryis given by \cite{PRs261-1} 
\begin{equation}
{\cal A}_{\parallel }=\frac 1m\left( 1-\epsilon \right) \left[ G_1\left(
E+E^{\prime }\cos \theta \right) -\frac{q^2}mG_2\right] \frac 1{W_1\left[
1+\epsilon R\right] }  \label{e122}
\end{equation}
For elastic scattering 
\begin{eqnarray}
\frac 1m\left( E+E^{\prime }\cos \theta \right) &=&2\left[ \frac Em-\tau -%
\frac{m\tau }E\right]  \nonumber \\
W_1 &=&G_M^2\delta \left( \nu -\frac{q^2}{2m}\right)  \label{e123} \\
R &=&\frac{\sigma _L}{\sigma _T}=\frac{G_E^2}{\tau G_M^2}  \nonumber \\
G_1 &=&\left( G_E+\tau G_M\right) \frac{G_M}{1+\tau }\delta \left( \nu -%
\frac{q^2}{2m}\right)  \nonumber \\
G_2 &=&-\frac 12\left( G_M-G_E\right) \frac{G_M}{1+\tau }\delta \left( \nu -%
\frac{q^2}{2m}\right)  \label{e124}
\end{eqnarray}
Hence we get from Eq. (\ref{e123}) 
\begin{equation}
{\cal A}_{\parallel }=2\left( 1-\epsilon \right) \frac \tau {1+\tau }\left( 
\frac{G_M}{\tau G_M^2+G_E^2}\right) \left\{ \left( \frac Em-\tau \frac mE%
\right) \left( G_E+\tau G_M\right) -\tau \left( 2G_E-\left( 1-\tau \right)
G_M\right) \right\}  \label{e125}
\end{equation}
Finally we discuss the polarization of the recoil proton For this case $%
n_\mu $ refers to recoil proton. Then, from Eq. (\ref{e93}), using Eqs. (\ref
{e71}--\ref{75}) and (\ref{e125}), we get \cite{PRL84-1398,PRL88-092301} 
\begin{eqnarray}
\left( \frac{d\sigma }{dq^2}\right) _T &=&\left( \frac{d\sigma }{dq^2}%
\right) _{\text{Mott}}\left[ -\frac{8\tau }{\sqrt{\tau \left( 1+\tau \right) 
}}\tan \frac \theta 2\right] G_MG_E  \label{e126} \\
\left( \frac{d\sigma }{dq^2}\right) _L &=&\left( \frac{d\sigma }{dq^2}%
\right) _{\text{Mott}}\left[ \frac{4\tau }{\sqrt{\tau \left( 1+\tau \right) }%
}\frac{E+E^{\prime }}m\tan ^2\frac \theta 2\right] G_M^2  \label{e127}
\end{eqnarray}
These equations can be put in the form 
\begin{eqnarray}
{\cal I}_T &=&\left( \frac{d\sigma }{dq^2}\right) _T/\left( \frac{d\sigma }{%
dq^2}\right)  \nonumber \\
&=&\frac{\epsilon \left( 1+\tau \right) }{\left( 1+\frac \epsilon \tau \frac{%
G_E^2}{G_M^2}\right) }\left[ -8\sqrt{\frac \tau {1+\tau }}\tan \frac \theta 2%
\right] \frac{G_E}{G_M}  \label{e128} \\
{\cal I}_L &=&\left( \frac{d\sigma }{dq^2}\right) _L/\left( \frac{d\sigma }{%
dq^2}\right)  \nonumber \\
&=&\frac{\epsilon \left( 1+\tau \right) }{G_M^2\left( 1+\frac \epsilon \tau 
\frac{G_E^2}{G_M^2}\right) }\left[ 4\sqrt{\frac \tau {1+\tau }}\frac{%
E+E^{\prime }}m\tan ^2\frac \theta 2\right] G_M^2  \label{e129} \\
\frac{G_E}{G_M} &=&-\left( \frac{{\cal I}_T}{{\cal I}_L}\right) \left( \frac{%
E+E^{\prime }}{2m}\right) \tan \frac \theta 2  \label{e130}
\end{eqnarray}
Equation (\ref{e130}) has been used in Refs. \cite{PRL84-1398,PRL88-092301}
to experimentally extract the $G_E/G_M$ by measuring the transverse and
longitudnal recoil proton polarization. Their result show a systematic
decrease of ratio $G_E/G_M$ as $q^2$ increases from 0.5 to 5.6 GeV$^2$,
indicating for the first time a definite difference in the spatial
distribution of charge and magnetization currents in the proton.

However, it is possible to extract the form factors $G_E$ and $G_M$ from the
longitudnal spin--spin asymmetry given in Eq. (\ref{e125}). From Eq. (\ref
{e125}), we can write 
\begin{eqnarray}
{\cal A}_{\parallel } &=&\frac{G_M}{\tau G_M^2+\epsilon G_E^2}\frac 1{%
1+\epsilon +2\epsilon \tau }\frac{2\tau }{1+\tau }\left\{ 2\epsilon \left(
1+\tau \right) \left( G_E+\tau G_M\right) \sqrt{\tau \left( 1+\tau \right)
\left( 1-\epsilon ^2\right) }\right.  \nonumber \\
&&+\left. \left( 1-\epsilon \right) \left[ 2G_E\left( \epsilon -\tau
+2\epsilon \tau ^2\right) +\tau G_M\left( 2\epsilon \left( 1+\tau \right)
+\left( 1-\tau \right) \left( 1+\epsilon +2\epsilon \tau \right) \right)
\right] \right\}  \label{e131}
\end{eqnarray}
Hence if we plot ${\cal A}_{\parallel }$ versus $\epsilon $ for fixed $\tau $%
, we can get information about the form factors $G_E$ and $G_M$. In
particular we note that for $\epsilon =0$%
\begin{eqnarray}
{\cal A}_{\parallel }\left( \epsilon =0\right) &=&\frac 2{G_M}\frac \tau {%
1+\tau }\left[ -2G_E+G_M-\tau G_M\right]  \nonumber \\
&=&\frac{2\tau }{1+\tau }\left[ 1-\tau -2\frac{G_E}{G_M}\right]  \nonumber \\
&=&2\tau \left[ \frac{F_2-F_1}{F_2+F_1}\right] =-2\tau \left[ \frac{1-F_2/F_1%
}{1+F_2/F_1}\right]  \label{e132}
\end{eqnarray}
This is an interesting result; it would supplement the result obtained from
the recoil proton polarization discussed above.

\section{Deep Inelastic Scattering}

We now briefly discuss the deep inelastic scattering \cite
{MIPP1992,PR162C45,PRD41-3517,PRs261-1,r20,r21}. In this region the
structure functions are found to be independent of four momentum transfer $%
q^2$ at fixed Bjorken variable $x=q^2/2m\nu $. In the scaling region both $%
q^2$ and $\nu $ are large but $x$ remains fixed. The deep inelastic
scattering is analysed in terms of Bjorken variable $x$ and inelasticity $y$%
. 
\begin{equation}
x=\frac{q^2}{2m\nu },\,\,\,\,\,y=\frac \nu E  \label{e133}
\end{equation}
It is convienent to introduce another variable 
\begin{equation}
\gamma ^2=\frac{q^2}{\nu ^2}=\frac{4m^2x^2}{q^2}  \label{e134}
\end{equation}
Interms of these variables we can write 
\begin{eqnarray}
\tan ^2\frac \theta 2 &=&\frac{\gamma ^2\left( 1-\epsilon \right) }{%
2\epsilon \left( 1+\gamma ^2\right) }=\frac{\gamma ^2y^2}{4\left( 1-y\right)
-\gamma ^2y^2}  \label{e135} \\
\epsilon &=&\frac{4\left( 1-y\right) -\gamma ^2y^2}{4\left( 1-y\right)
+2y^2+\gamma ^2y^2}  \label{e136}
\end{eqnarray}
We can express the electron--proton scattering cross-section given in Eq. (%
\ref{04}) in deep inelastic region as 
\begin{eqnarray}
\frac{d^2\sigma }{dq^2d\nu } &=&\left( \frac{d\sigma }{dq^2}\right) _{\text{%
Mott}}\frac{\gamma ^2F_1}{m\left( 1+\gamma ^2\right) \epsilon }\left[
1+\epsilon R\right]  \nonumber \\
\frac{d^2\sigma }{dxdy} &=&\left( \frac{d\sigma }{dq^2}\right) _{\text{Mott}%
}E\frac{4mxF_1}{\left( 1+\gamma ^2\right) \epsilon }\left[ 1+\epsilon
R\right]  \label{e137}
\end{eqnarray}
In deriving this result we have used 
\begin{eqnarray}
\nu W_2\left( \nu ,q^2\right) &=&F_2\left( x,q^2\right)  \nonumber \\
mW_1\left( \nu ,q^2\right) &=&F_1\left( x,q^2\right)  \label{e138} \\
\left( 1+\frac{\nu ^2}{q^2}\right) \frac{W_2}{W_1} &=&1+R  \nonumber \\
\frac{F_2\left( 1+\gamma ^2\right) }{2xF_1} &=&1+R  \label{e139} \\
2\left( 1+\frac{\nu ^2}{q^2}\right) \tan ^2\frac \theta 2 &=&\frac{%
1-\epsilon }{2\epsilon }  \label{e140}
\end{eqnarray}
The scattering cross-section for polarized longitudnal electrons on
longitudnally polarized protons given in Eq. (\ref{e117}) can be expressed
in the deep inelastic region as 
\begin{equation}
\frac{d^2\sigma ^{\overrightarrow{\leftarrow }}}{dxdy}=\left( \frac{d\sigma 
}{dq^2}\right) _{\text{Mott}}\frac{4mx\left( 1-\epsilon \right) }{\left(
1+\gamma ^2\right) \epsilon }\left[ g_1\left( 2-y-\frac 12\gamma
^2y^2\right) \frac 1y-\gamma ^2g_2\right]  \label{e141}
\end{equation}
where we have used $\nu G_1=g_1$, $\nu ^2G_2=mg_2$.

The longitudnal spin-spin asymmetry defined in Eq. (\ref{e122}) can be
written as 
\begin{equation}
{\cal A}_{\parallel }=\left( 1-\epsilon \right) \frac 1{F_1\left( 1+\epsilon
R\right) }\left[ g_1\left( 2-y-\frac 12\gamma ^2y^2\right) \frac 1y-\gamma
^2g_2\right]  \label{e142}
\end{equation}
Since $\gamma ^2\rightarrow 0$ as $q^2\rightarrow \propto $; for large $q^2$%
, it is a good approximation to put $\gamma ^2=0$. In what follows we will
make this approximation. Thus we take 
\begin{eqnarray}
\epsilon &\approx &\frac{4\left( 1-y\right) }{4\left( 1-y\right) +2y^2}=%
\frac{2\left( 1-y\right) }{1+\left( 1-y\right) ^2}\equiv \frac{2\left(
1-y\right) }{Y_{+}}  \nonumber \\
1-\epsilon &\approx &\frac{y^2}{Y_{+}}  \nonumber \\
1-\epsilon \left( 1-y\right) &=&\frac{1-\left( 1-y\right) ^2}{1+\left(
1-y\right) ^2}=\frac{Y_{-}}{Y_{+}}=\left( 1-\epsilon \right) \left( \frac{2-y%
}y\right)  \label{e143}
\end{eqnarray}
Thus we get a simple expression for the longitudnal spin-spin asymmetry \cite
{PRs261-1} 
\begin{equation}
{\cal A}_{\parallel }=\frac{g_1\left[ 1-\epsilon \left( 1-y\right) \right] }{%
F_1\left( 1+\epsilon R\right) }  \label{e144}
\end{equation}
This gives 
\begin{equation}
g_1=\frac{{\cal A}_{\parallel }}D\frac{\left( 1+\gamma ^2\right) F_2}{%
2x\left( 1+R\right) },\,\,\,\,\,D=\frac{1-\epsilon \left( 1-y\right) }{%
1+\epsilon R}  \label{e145}
\end{equation}
Similarly the parity violating scattering cross-section of electrons on
longitudnal polarized protons given in Eq. (\ref{65})can be expressed in
deep inelastic region \cite{PRD38-3390} 
\begin{eqnarray}
\frac{d^2\sigma ^{\rightarrow }}{dq^2d\nu } &=&\frac{G_Fq^2}{\sqrt{2}\pi
\alpha }4mE\frac{1-\epsilon }{4\epsilon }\left( \frac{d\sigma }{dq^2}\right)
_{\text{Mott}}  \nonumber \\
&&\times \left\{ v_e\left[ \frac{2\left( 1-y\right) }{y^2}\tilde{h}_2^e-x%
\tilde{h}_3^e-\left( 1-y\right) \frac 1{y^2}\tilde{h}_4^e\right] +a_e\left[ 
\tilde{g}_1^e\frac{2-y}y\right] \right\}  \label{e146}
\end{eqnarray}
where we have used, 
\begin{equation}
\nu \tilde{G}_1^e=\tilde{g}_1^e,\text{ }\nu ^2\tilde{G}_2^e=m\tilde{g}_2^e,%
\text{ }\nu \tilde{H}_2^e=\tilde{h}_2^e,\text{ }\nu \tilde{H}_3^e=\tilde{h}%
_3^e,\text{ and }\nu \tilde{H}_4^e=m\tilde{h}_4^e.  \label{e147}
\end{equation}
Now using the light cone algebra sum rules \cite{PRD38-3390} 
\begin{eqnarray}
\tilde{h}_3^e\left( x\right) &=&-\frac 1x\tilde{h}_2^e\left( x\right) 
\nonumber \\
\tilde{h}_4^e\left( x\right) &=&0  \label{e148}
\end{eqnarray}
we get 
\begin{eqnarray}
\frac{d^2\sigma ^{\rightarrow }}{dxdy} &=&\frac{G_Fq^2}{\sqrt{2}\pi \alpha }%
4mE\frac{1-\epsilon }{4\epsilon }\left( \frac{d\sigma }{dq^2}\right) _{\text{%
Mott}}\frac 1{y^2}  \nonumber \\
&&\times \left\{ v_e\left[ 2\left( 1-y\right) +y^2\right] \tilde{h}%
_2^e+a_e\left[ \tilde{g}_1^exy\left( 2-y\right) \right] \right\}
\label{e149}
\end{eqnarray}
Hence the longitudnal proton asymmetry is given by 
\begin{eqnarray}
{\cal A}_L^p &=&\frac{\frac{d^2\sigma ^{\leftarrow }}{dxdy}-\frac{d^2\sigma
^{\rightarrow }}{dxdy}}{\frac{d^2\sigma ^{\leftarrow }}{dxdy}+\frac{%
d^2\sigma ^{\rightarrow }}{dxdy}}  \nonumber \\
&=&-\frac{G_Fq^2}{\sqrt{2}\pi \alpha }\left( \frac{1-\epsilon }4\right) 
\frac{v_eY_{+}\frac 1xh_2^e\left( x,q^2\right) +a_eY_{-}\tilde{g}_1^e\left(
x,q^2\right) }{F_1\left( 1+\epsilon R\right) y^2}  \nonumber \\
&=&-\frac{G_Fq^2}{\sqrt{2}\pi \alpha }\frac{v_e\frac 1xh_2^e\left(
x,q^2\right) +a_e\left[ 1-\epsilon \left( 1-y\right) \right] \tilde{g}%
_1^e\left( x,q^2\right) }{4F_1\left( 1+\epsilon R\right) }  \label{e150}
\end{eqnarray}

Finally for the parity violating polarized electron scattering on
unpolarized proton, we get [cf. Eq. (\ref{86})] 
\begin{eqnarray}
\frac{d^2\sigma ^{\rightarrow }}{dxdy} &=&\frac{G_Fq^2}{\sqrt{2}\pi \alpha }%
\left( \frac{d\sigma }{dq^2}\right) _{\text{Mott}}  \nonumber \\
&&\times \left\{ v_e\frac{Y_{-}}{Y_{+}}x\tilde{F}_3-a_e\left[ \tilde{F}_2-%
\frac{y^2}{Y_{+}}\tilde{F}_L\right] \right\}  \label{e151}
\end{eqnarray}
where we have used the relation 
\begin{equation}
2x\tilde{F}_1=\tilde{F}_2-\tilde{F}_L  \label{e152}
\end{equation}
Hence the longitudnal electron asymmetry 
\begin{eqnarray}
{\cal A}^e &=&\frac{G_Fq^2}{\sqrt{2}\pi \alpha }\frac{v_e\frac{Y_{-}}{Y_{+}}x%
\tilde{F}_3-a_e\left[ \tilde{F}_2-\frac{y^2}{Y_{+}}\tilde{F}_L\right] }{%
2xF_1\left( 1+\epsilon R\right) }  \nonumber \\
&=&\frac{G_Fq^2}{\sqrt{2}\pi \alpha }\frac{v_e\left[ 1-\epsilon \left(
1-y\right) \right] x\tilde{F}_3-a_e\left[ \tilde{F}_2-\left( 1-\epsilon
\right) \tilde{F}_L\right] }{2xF_1\left( 1+\epsilon R\right) }  \label{e153}
\end{eqnarray}
>From the experimental measurements of asymmetries ${\cal A}_{\Vert }$ and $%
{\cal A}_L^p$, the structure functions $g_1$, $\tilde{g}_1^e$ and $\tilde{h}%
_2^e$ can be extracted. Note that they are functions of $x$; their
dependence on $q^2$ is weak (logrithmic) and can be taken care of by QCD
corrections. These structure functions in turn give information about the
fundamental constituents of the proton, prticularly their spin content. The
asymmetry ${\cal A}_L^p$ involve two functions $\tilde{g}_1^e$ and $\tilde{h}%
_2^e$. The structure function $\tilde{g}_1^e$ is multiplied by a y-dependent
term, whereas $\tilde{h}_2^e$ is just multiplied by a constant. Hence by
measuring ${\cal A}_L^p$ at various values of $y$, it is possible to extract
both $\tilde{h}_2^e$ and $\tilde{g}_1^e$.

The spin-dependent structure functions $g_1$ and $\tilde{g}^e$ satisfy the
sum rules given below; the right-hand-side of these sum rules is given in
terms of the spin-content of elementry constituents of proton.

For the structure function $g_1\left( x\right) $, light cone algebra gives
the sum rule 
\begin{equation}
\int_0^1g_1^{p,n}\left( x\right) dx=\left[ \pm \frac 16a_3+\frac 1{6\sqrt{3}}%
a_8+\frac 13\sqrt{\frac 23}a_0\right]  \label{e154}
\end{equation}
$a_3$ and $a_8$ are given as follows 
\begin{eqnarray}
a_3 &=&\frac 12g_A=\frac 12\left( F+D\right)  \nonumber \\
a_8 &=&\frac 1{2\sqrt{3}}g_A=\frac 1{2\sqrt{3}}\left( 3F-D\right)
\label{e155}
\end{eqnarray}
The $a_0$ being singlet cannot be written in terms of $F$ and $D$; but we
put 
\begin{equation}
a_0=\frac 12\sqrt{\frac 23}g_A^0  \label{e156}
\end{equation}
Hence, from Eq. (\ref{e154}), using Eqs. (\ref{e155}) and (\ref{e156}), we
get the Bjorken sum rule \cite{PRD14-1467} 
\begin{equation}
\int_0^1\left[ g_1^p\left( x\right) -g_1^n\left( x\right) \right] dx=\frac 16%
g_A\left( 1-\frac{\alpha _s\left( q^2\right) }\pi \right)  \label{e157}
\end{equation}
The factor multiplying $g_A$ is the QCD correction to the sum rule. Bjorken
sum rule is well satisfied experimentally. Also for proton only, the sum
rule can be written as follows 
\begin{equation}
\int_0^1g_1^p\left( x\right) dx=\frac 1{12}\left[ \left( F+D\right) +\frac 13%
\left( 3F-D\right) +\frac 43g_A^0\right]  \label{e158}
\end{equation}
However, if we take 
\begin{equation}
g_A^0=g_8=3F-D,  \label{e159}
\end{equation}
we get the Ellis--Jafee sum rule \cite{PRD9-1444} 
\begin{equation}
\int_0^1g_1^p\left( x\right) dx=\frac 1{12}\left[ 1+\frac 53\frac{3F-D}{F+D}%
\right]  \label{e160}
\end{equation}
This sum rule is in disagreement with the data. This shows that Eq. (\ref
{e159}) does not hold experimentally viz the strange content of the proton
is not zero.

In the naive quark--parton model, the sum rule (\ref{e159}) can be written
as 
\begin{equation}
\int_0^1g_1^{p,n}\left( x\right) dx=\frac 12\sum_qe_q^2\Delta q  \label{e161}
\end{equation}
where 
\begin{equation}
\Delta q=\int_0^1\left\{ \left[ q^{\uparrow }\left( x\right) +\bar{q}%
^{\uparrow }\left( x\right) \right] -\left[ q^{\downarrow }\left( x\right) +%
\bar{q}^{\downarrow }\left( x\right) \right] \right\}  \label{e162}
\end{equation}
Here $\Delta q$ is the quark contribution to the first moment of $g_1\left(
x\right) $. There is also gluon contribution to it. To include this we
replace $\Delta q$ by $\Delta \tilde{q}$ defined in Eqs. (\ref{36}) and (\ref
{37}). Then using Eqs. (38), we can write the sum rules (\ref{e157}) and (%
\ref{e158}) as (same results follow from Eq. (\ref{e161})) 
\begin{eqnarray}
\int_0^1\left[ g_1^p\left( x\right) -g_1^n\left( x\right) \right] dx &=&%
\frac 16\left( \Delta \tilde{u}-\Delta \tilde{d}\right) \left( 1-\frac{%
\alpha _s\left( q^2\right) }\pi \right)  \label{e163} \\
\int_0^1g_1^p\left( x\right) dx &=&\frac 1{12}\left[ \left( \Delta \tilde{u}%
-\Delta \tilde{d}\right) +\frac 13\left( \Delta \tilde{u}-\Delta \tilde{d}%
-2\Delta \tilde{s}\right) +\frac 43\left( \Delta \tilde{u}+\Delta \tilde{d}%
+\Delta \tilde{s}\right) \right]  \nonumber \\
&=&\frac 12\left[ \frac 49\Delta \tilde{u}+\frac 19\Delta \tilde{d}+\frac 19%
\Delta \tilde{s}\right]  \label{e164}
\end{eqnarray}
Now we discuss the sum rule for $\tilde{g}_1^{ep}$. Light cone algebra gives
the sum rule 
\begin{eqnarray}
\int_0^1\left[ \tilde{g}_1^{ep}\left( x\right) +2x_Wg_1^p\left( x\right)
\right] dx &=&\left[ \frac 1{12}a_3+\frac 1{4\sqrt{3}}a_8+\frac 14\sqrt{%
\frac 23}a_0\right]  \nonumber \\
&=&\frac 1{24}\left[ \left( F+D\right) +\left( 3F-D\right) +2g_A^0\right] 
\nonumber \\
&=&\frac 1{12}\left[ 2\Delta \tilde{u}+\Delta \tilde{d}\right]  \label{e165}
\end{eqnarray}
However, from Eqs. (\ref{e165}) and (\ref{e158}), eleminating the singlet
contribution, we get the sum rule 
\begin{eqnarray}
\int_0^1\left[ 4\tilde{g}_1^{ep}\left( x\right) -\left( 3-8x_W\right)
g_1^p\left( x\right) \right] dx &=&\frac 16\left( F-D\right) \left( 1-\frac{%
\alpha _s\left( q^2\right) }\pi \right)  \nonumber \\
&=&\frac 16\left( \Delta \tilde{d}-\Delta \tilde{s}\right) \left( 1-\frac{%
\alpha _s\left( q^2\right) }\pi \right)  \label{e166}
\end{eqnarray}
Note that this sum rule involves only the proton target.

We conclude that the sum rules given in Eqs. (\ref{e163}) and (\ref{e166})
are independent of unknown signet axial vector constant $g_A^0$.

The sum rules (\ref{e163}), (\ref{e164}), and (\ref{e165}) provide means to
extract $\Delta \tilde{u}$, $\Delta \tilde{d}$, and $\Delta \tilde{s}$ from
deep inelastic scattering without any put from the hypron $\beta $-decay.
This would test the consistency of electro-weak theory at low energy with
that in the deep inelastic region. This will be possible only in the future
experiments with the measurement of ${\cal A}_L^p$.

Finally it is convienient to write the asymmetry ${\cal A}^e$ given in Eq. (%
\ref{e153}) in terms of the following structure functions. 
\begin{eqnarray}
F_2 &=&\frac 12\left( F_2^{ep}+F_2^{en}\right)  \nonumber \\
\tilde{F}_2 &=&\frac 12\left( \tilde{F}_2^{ep}+\tilde{F}_2^{en}\right) 
\nonumber \\
\tilde{F}_3 &=&\frac 12\left( \tilde{F}_3^{ep}+\tilde{F}_3^{en}\right) 
\nonumber \\
F_2^{cc} &=&\frac 12\left( F_2^{\bar{\nu}p}+F_2^{\nu p}\right)  \nonumber \\
F_3^{cc} &=&\frac 12\left( F_3^{\bar{\nu}p}+F_3^{\nu p}\right)  \label{e168}
\end{eqnarray}
The following relations hold between these structure functions 
\begin{eqnarray}
\tilde{F}_2 &=&\frac 14F_2^{cc}-2x_WF_2  \nonumber \\
\tilde{F}_3 &=&\frac 14F_3^{cc}  \nonumber \\
F_2^{cc}-\frac{18}5F_2 &=&4x\frac 15\left[ \frac 2{\sqrt{3}}v_8-\sqrt{\frac 2%
3}v_0\right]  \nonumber \\
\tilde{F}_2 &=&4x\frac 1{12}\left[ \left( 3-4x_W\right) \frac 1{\sqrt{3}}%
v_8+\left( 3-8x_W\right) \sqrt{\frac 23}v_0\right]  \label{e169}
\end{eqnarray}
If 
\begin{equation}
v_0=\sqrt{2}v_8,  \label{e170}
\end{equation}
then 
\begin{eqnarray}
\frac{F_2}{\frac 5{18}F_2^{cc}} &=&1\,\,\,\left( 1.007\pm 0.063\right)
\label{e171} \\
\frac{\tilde{F}_2}{F_2} &=&\frac 9{10}\left( 1-\frac{20}9x_W\right)
\label{e172}
\end{eqnarray}
The experimental value given in Eq. (\ref{e171}) shows that the condition (%
\ref{e170}) is well satisfied experimentally. It verifies that charges of $u$
and $d$ valence quarks as their mean charges is 5/18 i.e. there is no sea of
strange quarks. From the experimental measurement of asymmetry ${\cal A}^e$
given in Eq. (\ref{e153}) at various values of $y$, $\tilde{F}_2$ can be
extracted as the leading contribution comes from the term containing $\tilde{%
F}_2$ which is multiplied by a constant $a_e=-1$ independent of $y$. It
would test Eq. (\ref{e172}) experimentally.

The electron proton scattering is a very direct means of probing the
structure of proton. We have given a unified approach for the elastic and
the highly inelastic scattering of electrons of proton target.In the case of
elastic scattering, the structure functions reduce to the form factors which
are functions of $q^2$. These form factors give the spatial distribution of
charge and magnetization currents in the proton. The Rosenbluth extraction
of the electromagnetic form factors $G_E$ and $G_M$ has been supplemented
recently by measurements of recoil polarization of proton. These experiments
indicate break down of electric and magnetic distribution at high $q^2$ in a
proton. In this respect, we have pointed out in this article that the
longitudnal spin-spin asymmetry ${\cal A}_{\Vert }$ in the elastic
scattering of polarized electrons of polarized protons may provide an
additional probe of these form factors. In particular the asymmetry ${\cal A}%
_{\Vert }$ at $\epsilon =0$, would directly give the ratio $F_2/F_1$.The
parity violating probe of the proton by scattering of polarized
(unpolarized) electrons on unpolarized (polarized) protons would give
additional information about the structure of proton both in the elastic and
highly inelastic region. In this respect the measurements of the asymmetries 
${\cal A}_L^p$ and ${\cal A}^e$ are of particular importance for the strange
content of the proton. Moreover we have pointed out that the observation of
recoil polarization of proton induced by electro-weak interference in the
elastic scattering of electrons on proton target will be of interest for the
structure of proton.

The author would like to thank Amjad Gilani for drawing the figures.

Figure Captions

\begin{enumerate}
\item  Scattering of electrons on nucleons \label{fig1}

\item  Plot of the longitudnal asymmetry ${\cal A}_L^p$ versus $\epsilon $
[see Eq. (\ref{e103})] for various values of $q^2$; Upper and lower figures
correspond to the sets of parameters (i) and (ii) given in Eq. (\ref{e98})
and (\ref{e99}) respectively.\label{fig2} Solid line for $q^2=0.5$, dashed
line $q^2=1$, dotted line $q^2=2.64$, dash-dotted line $q^2=3.2$,
dash-dot-dot line $q^2=5.6$, dash-dot-dot-dot line $q^2=10$.

\item  Plot of the longitudnal asymmetry ${\cal A}^e$ versus $\epsilon $
[see Eq. (\ref{e104})], line description is same as Fig. \ref{fig2}.

\item  Plot of the longitudnal recoil polarization of proton $I_L$ versus $%
\epsilon $ [see Eq. (\ref{e105})], line description is same as Fig. \ref
{fig2}.

\item  Plot of the transverse recoil polarization of proton $I_T$ versus $%
\epsilon $ [see Eq. (\ref{e106})], line description is same as Fig. \ref
{fig2}.
\end{enumerate}

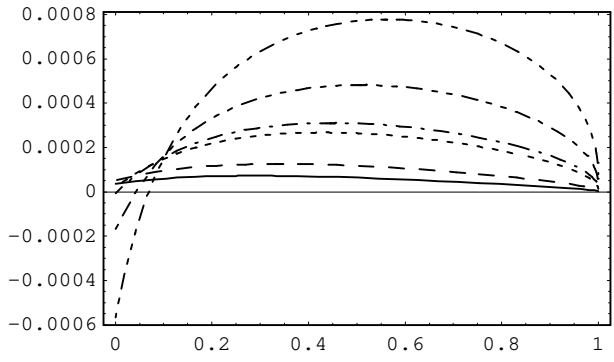
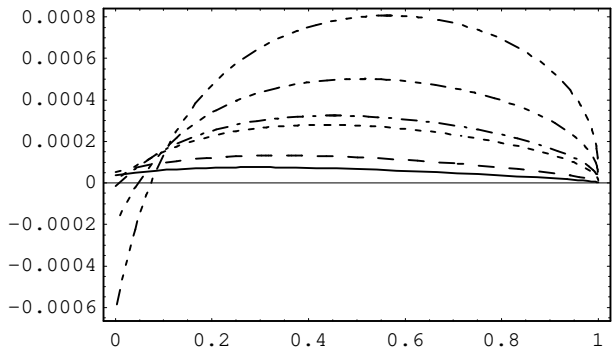
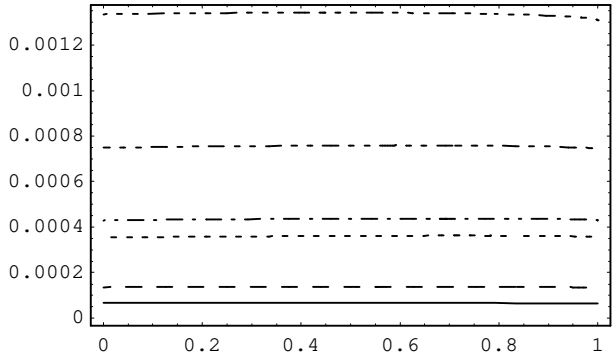
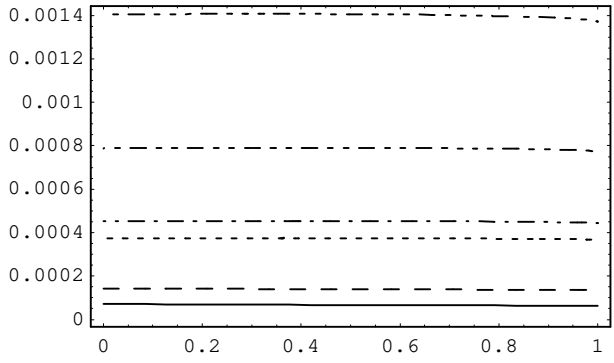
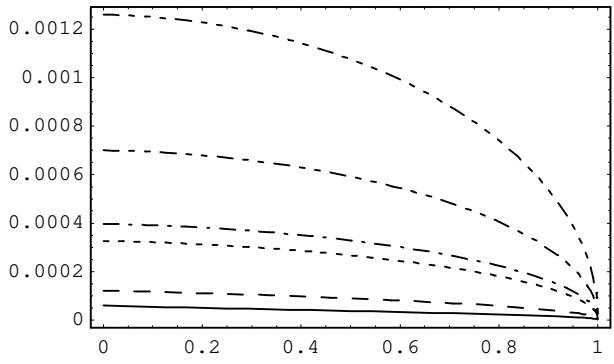
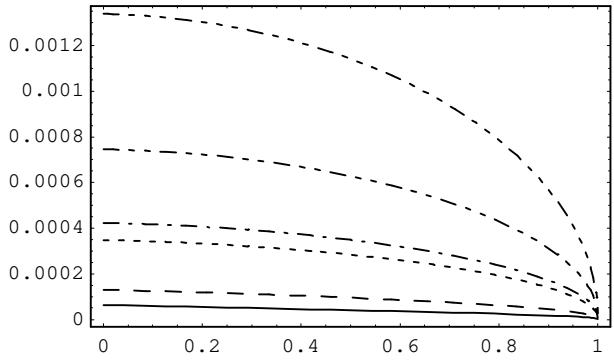
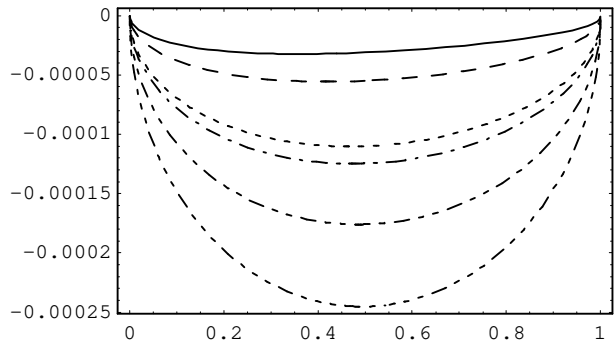
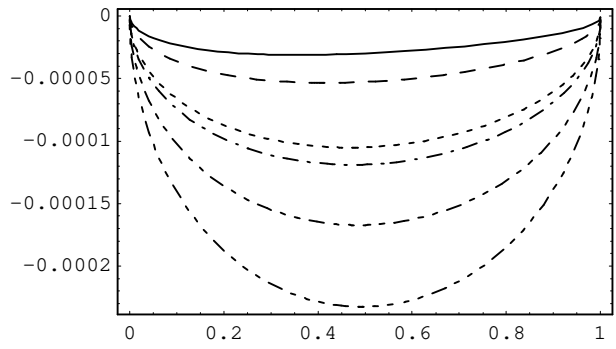
\end{document}